\documentclass[fleqn,usenatbib]{mnras}

\usepackage{newtxtext,newtxmath}

\usepackage[T1]{fontenc}
\usepackage{xcolor, color}

\DeclareRobustCommand{\VAN}[3]{#2}
\let\VANthebibliography\thebibliography
\def\thebibliography{\DeclareRobustCommand{\VAN}[3]{##3}\VANthebibliography}

\usepackage{graphicx}	
\usepackage{amsmath}	
\usepackage{amssymb}	
\usepackage{bm}

\newcommand{\Omegac}{\Omega_{\rm c}}
\newcommand{\Ic}{I_{\rm c}}
\newcommand{\Nc}{N_{\rm c}}
\newcommand{\xic}{\xi_{\rm c}}
\newcommand{\tauc}{\tau_{\rm c}}
\newcommand{\sigmac}{\sigma_{\rm c}}
\newcommand{\sigcic}{\frac{\sigma_{\rm c}^2}{\Ic^2}}

\newcommand{\Omegas}{\Omega_{\rm s}}
\newcommand{\Is}{I_{\rm s}}
\newcommand{\Ns}{N_{\rm s}}
\newcommand{\xis}{\xi_{\rm s}}
\newcommand{\taus}{\tau_{\rm s}}
\newcommand{\sigmas}{\sigma_{\rm s}}
\newcommand{\sigsis}{\frac{\sigma_{\rm s}^2}{\Is^2}}

\title[Parameter estimation of a neutron star model]{Rapid parameter estimation of a two-component neutron star model with spin wandering using a Kalman filter}

\author[P. M. Meyers et al.]{
Patrick M. Meyers,$^{1,2}$\thanks{E-mail: pat.meyers@unimelb.edu.au}
Nicholas J. O'Neill,$^{1}$
Andrew Melatos$^{1,2}$,
Robin J. Evans$^{3,2}$
\\
$^{1}$School of Physics, University of Melbourne, Parkville, VIC 3010, Australia\\
$^{2}$OzGrav, University of Melbourne, Parkville, VIC 3010, Australia\\
$^{3}$Department of Electrical and Electronic Engineering, University of Melbourne, Parkville, Victoria 3010, Australia
}

\date{Accepted XXX. Received YYY; in original form ZZZ}

\pubyear{2021}

\begin{document}
\label{firstpage}
\pagerange{\pageref{firstpage}--\pageref{lastpage}}
\maketitle

\begin{abstract}
The classic, two-component, crust-superfluid model of a neutron star can be formulated as a noise-driven, linear dynamical system, in which the angular velocities of the crust and superfluid are tracked using a Kalman filter applied to electromagnetic pulse timing data and gravitational wave data, when available. Here it is shown how to combine the marginal likelihood of the Kalman filter and nested sampling to estimate full posterior distributions of the six model parameters, extending previous analyses based on a maximum-likelihood approach. The method is tested across an astrophysically plausible parameter domain using Monte Carlo simulations. It recovers the injected parameters to $\lesssim 10$ per cent for time series containing $\sim 10^3$ samples, typical of long-term pulsar timing campaigns. It runs efficiently in $\mathcal O(1)$ CPU-hr for data sets of the above size. In a present-day observational scenario, when electromagnetic data are available only, the method accurately estimates three parameters: the relaxation time, the ensemble-averaged spin-down of the system, and the amplitude of the stochastic torques applied to the crust. In a future observational scenario, where gravitational wave data are also available, the method also estimates the ratio between the moments of inertia of the crust and the superfluid, the amplitude of the stochastic torque applied to the superfluid, and the crust-superfluid lag. These empirical results are consistent with a formal identifiability analysis of the linear dynamical system.
\end{abstract}




\section{Introduction}

Rotating neutron stars are ideal candidates for multimessenger experiments of bulk nuclear matter~\citep{2017IJMPD..2630015G,2018ASSL..457..673G}. Precision pulsar timing experiments in both the X-ray and radio bands continue to yield ground-breaking measurements and discoveries. Most recently, this includes tight constraints on the mass and radius of neutron stars using the Neutron-star Interior Composition Explorer~\citep{10.1117/12.2231304,Riley_2019,Raaijmakers_2019,Miller_2019,Bogdanov_2019}, as well as evidence for a common noise process in the NANOGrav 12.5 year data set that is consistent with the shape of a stochastic gravitational-wave background~\citep{2020ApJ...905L..34A}.  One common feature revealed by pulsar timing experiments is deviations from the long-term secular rotation and spin-down of the star known as `spin-wandering.' When testing general relativity or seeking to detect a stochastic gravitational--wave background, spin wandering is treated as a nuisance, which needs to be characterized accurately and then subtracted in a statistical sense~\citep{1975ApJS...29..453G,1980ApJ...237..216C,1994ApJ...422..671A,Shannon2010,2012MNRAS.426.2507P,2019MNRAS.487.5854N,2019MNRAS.489.3810P,2020MNRAS.494..228L,parthasarathy:2020wel,2020MNRAS.497.3264G}. In this paper, we use `spin wandering' to refer to the achromatic fluctuations in pulse times-of-arrival (TOAs) that are intrinsic to the pulsar, specifically the rotation of its crust and corotating magnetosphere. This is in contrast to chromatic TOA fluctuations caused by propagation through the magnetosphere and interstellar medium~\citep{2013MNRAS.429.2161K,2013CQGra..30v4002C,2014ApJ...790L..22A,2016ApJ...818..166L,2016MNRAS.458.2161L,2017ApJ...834...35L,2020arXiv200810562D,2021MNRAS.502..478G}. Together the achromatic and chromatic TOA fluctuations are usually termed `timing noise'.

Stochastic torques can excite deterministic dynamical modes in the star, such as relaxation processes. Thus the specific, random realization of the spin wandering we observe in a neutron star contains useful information about the star's structure~\citep{Baykal:1991uec,2012MNRAS.426.2507P,Melatos2014} or external influences on the star~\citep{1997ApJS..113..367B,Mukherjee:2018rim}. In the popular two-component, crust-superfluid model~\citep{1969Natur.224..872B}, a stochastic driving torque acting on the crust is counteracted by the restoring torque from the coupling between the crust and the superfluid. \cite{exp_max_paper_meyers_melatos} showed with synthetic data that one can resolve the natural relaxation time-scale of the two-component system by using a Kalman filter to track the spin wandering of the crust without seeking to subtract or average out the ``noisy'' component of the signal. The relaxation time-scale has also been estimated by auto-correlating noise residuals~\citep{2012MNRAS.426.2507P}.

\cite{exp_max_paper_meyers_melatos} presented a maximum-likelihood method to solve for the parameters of the two-component model based on the expectation-maximization algorithm~\citep{em_original_paper,Shumway1982, GIBSON20051667}. They considered two observational scenarios. In the first, measurements of the rotation period of the crust (electromagnetic observations) and the core (gravitational-wave observations) are available. In the second, only measurements of the rotation period of the crust are available. Using synthetic data, one can show that in the first observational scenario it is possible to accurately estimate all six of the two-component model parameters. In the second observational scenario we are only able to accurately estimate two out of the six parameters. 

The framework in~\cite{exp_max_paper_meyers_melatos} is nearly ready to be implemented as an analysis pipeline for real astronomical data. However, it can be improved in three important ways, which are developed in this paper. First, in certain applications it is instructive physically to map out the posterior probability density of the model parameters, instead of focusing on their modal values as in the maximum-likelihood approach. Here we use a Markov chain Monte Carlo (MCMC) sampler to achieve this goal. Second,~\cite{exp_max_paper_meyers_melatos} made a first-order Euler approximation to simplify the state transitions, angular impulse, and process-noise variance during a time step. The Euler approximation limits the accuracy with which the parameters can be estimated in some circumstances. Here we replace it with the full, analytic solution of the stochastic differential equations that define the two-component model, eliminating the artificial need to sample the data faster than the dynamical time-scale of the system. This is a major practical improvement for radio timing experiments, where times-of-arrival are measured typically over days to weeks instead of hours. Third, we clarify from first principles which parameters can be estimated reliably under both the electromagnetic-gravitational and electromagnetic-only observational scenarios, and which parameters are inaccessible irrespective of the size and quality of the data set. This exercise is known as assessing ``identifiability''~\citep{BELLMAN1970329}. The model in this paper is a physically-motivated special case of the abstract class of autoregressive moving average models for an arbitrary red-noise process formulated by~\cite{Kelly_2014}.

The rest of this paper is organized as follows. In Section~\ref{sec:two_component_model} we present the two component model of the neutron star of~\cite{1969Natur.224..872B}, discuss its analytic solution, and formulate it as a hidden Markov model. We then present an analysis of identifiability in the two observational scenarios above. In Section~\ref{sec:parameter_estimation} we briefly discuss Bayesian parameter estimation before defining the Kalman filter likelihood function. In Section~\ref{sec:results} we apply this new method to synthetic data to estimate the posterior distribution of the model parameters under the two observational scenarios. We interpret the results in the context of identifiability. In Section~\ref{sec:conclusion} we discuss the implications for future analyses with real astronomical data and additional potential refinements of the method.

\section{Two-component neutron star}
\label{sec:two_component_model}

The two-component model for a rotating neutron star was originally motivated by the post-glitch recovery seen in the Vela pulsar~\citep{1969Natur.224..872B}. It comprises a superfluid core, believed to be an inviscid neutron condensate, and a crystalline crust which locks magnetically to the charged fluid species (either electrons or superconducting protons) in the inner crust and outer core~\citep{Mendell1991a,Mendell1991,2006CQGra..23.5505A,2011MNRAS.410..805G}. Both components are assumed to rotate uniformly for simplicity. Their angular velocities are unequal in general.

We introduce and briefly discuss the coupled equations of motion that the two components obey in Section~\ref{ssec:two_component_model:equations_of_motion}. In Section~\ref{ssec:two_component_model:state_space_representation} we present a state-space representation of the model, which tracks the wandering angular velocity of each component. The state-space representation discretizes the two-component dynamics in a form which feeds neatly into the parameter estimation framework developed in Section~\ref{sec:parameter_estimation}, which uses the Kalman filter to evaluate the likelihood used in an MCMC sampler. In Section~\ref{ssec:two_component_model:identifiability}, we discuss which parameters can be estimated accurately when we have at our disposal measurements of the angular velocity of the crust and core~(Section~\ref{sssec:identifiability:em_gw_observations}) and the crust only~(Section~\ref{sssec:identifiability:em_only_observations}).

\subsection{Equations of motion}
\label{ssec:two_component_model:equations_of_motion}
The stellar components obey the coupled equations of motion
\begin{align}
    \Ic\frac{\textrm{d}\Omegac}{\textrm{d}t} &= \Nc + \xic(t) - \frac{\Ic}{\tauc}\left(\Omegac - \Omegas\right)\label{eq:crust_equation_of_motion},\\
    \Is\frac{\textrm{d}\Omegas}{\textrm{d}t} &= \Ns + \xis(t) - \frac{\Is}{\taus}\left(\Omegas - \Omegac\right)\label{eq:superfluid_equation_of_motion},
\end{align}
where the subscripts ``c'' and ``s'' label the crust and the superfluid respectively, $\Omegac$ and $\Omegas$ are angular velocities, $\Ic$ and $\Is$ are effective moments of inertia, $\Nc$ and $\Ns$ are secular torques, $\xic$ and $\xis$ are zero-mean stochastic torques, and $\tauc$ and $\taus$ are coupling time-scales.

The model is highly idealized. For example, uniform rotation within each component breaks down for $\Omegac\neq\Omegas$~\citep{Reisenegger:1993aab,Abney:1996aaaa,Peralta2005,2010MNRAS.409.1253V,van_Eysden_2013}, so the quantities $\Ic$, $\Is$, $\Omegac$, $\Omegas$ represent body-averaged approximations to the realistic behavior~\citep{2010MNRAS.405.1061S,haskell:2012rds}. Moreover, the pinning of the superfluid to nuclear lattice sites leads to nonlinear stick-slip dynamics in $\Omegas$ which do not emerge explicitly from equation (\ref{eq:superfluid_equation_of_motion})~\citep{Warszawski:2011ldo,melatos:2013pwe,drummond:2017wfg,drummond:2018esf,2018PASA...35...20K,2019MNRAS.487..702L}. Finally, equations (\ref{eq:crust_equation_of_motion}) and (\ref{eq:superfluid_equation_of_motion}) do not consider magnetohydrodynamic forces~\citep{2011MNRAS.410..805G}. Such forces are typically important dynamically with respect to the small lag $|\Omega_{\rm s} - \Omega_{\rm c}| \ll \Omegac,\Omegas$. The steady-state coexistence of differential rotation and internal magnetic fields with open and closed topologies is a subtle problem which has been studied analytically \citep{Easson:1979edf,melatos:2012ped,glampedakis:2015yhf} and numerically \citep{2020MNRAS.494.3095A,2020MNRAS.495.1360S,2021arXiv210414908S}. The time-dependent interaction between differential rotation and the Lorentz force has also been studied in the context of magnetized Couette flows and instabilities \citep{2019PhRvF...4j3905M,2020arXiv200810921R}. The MHD equations of motion for a magnetized, differentially rotating, and possibly superconducting Fermi liquid, relevant to neutron stars, have been derived within a multicomponent framework by several authors \citep{Easson:1979edf,Mendell1991a,Mendell1991,1998MNRAS.296..903M,2011MNRAS.410..805G,2012MNRAS.420.1263G,2012MNRAS.424..482L,2018ApJ...865...60V}.

The interpretations of $\Nc$ and $\Ns$ depend on whether the neutron star is accreting or isolated. The crust, for example, corotates with the large-scale stellar magnetic field and experiences a magnetic dipole braking torque~\citep{goldreich:1969wiu}. It also experiences a gravitational radiation reaction torque if it has a thermally or magnetically induced mass quadrupole moment~\citep{ushomirsky:2000bvc,melatos:2005yup}. These two scenarios imply $\Nc < 0$. If the crust experiences a hydromagnetic accretion torque as well, then one has either $\Nc > 0$ or $\Nc < 0$~\citep{Ghosh1979,1997ApJS..113..367B,romanova2004khj}. The superfluid, meanwhile, is decoupled electromagnetically from the crust and should not be directly influenced by dipole braking or accretion\footnote{Pinning to quantized magnetic flux tubes in a type II superconductor changes this picture, introducing glassy dynamics into the superfluid response~\citep{drummond:2017wfg,drummond:2018esf}.}. However, it may have a time-varying mass or current quadrupole moment that results in a gravitational radiation reaction torque~\citep{alpar:1996aew,sedrakian:2002vbf,jones:2006aok,2019AIPC.2127b0007H,melatos:2010bvy,Melatos2014}. In this paper $\Nc$ and $\Ns$ are treated as constant for the sake of simplicity. In general one has $\Nc\propto\Omegac^3$ for magnetic dipole braking~\citep{1967Natur.216..567P,1969Natur.221..454G} and $N_{\rm c,s}\propto\Omega_{\rm c,s}^5$ for gravitational radiation reaction~\citep{1969ApJ...158L..71F}.

The right-most terms of equations~(\ref{eq:crust_equation_of_motion}) and (\ref{eq:superfluid_equation_of_motion}) couple the crust and the superfluid. They reduce the lag, $|\Omegac - \Omegas|$, between the components and form an action-reaction pair when $\Ic/\tauc=\Is/\taus$. This coupling can arise physically through vortex-mediated mutual friction~\citep{1969Natur.224..872B, Mendell1991} and entrainment~\citep{1976JETP...42..164A,2006CQGra..23.5505A}. Here we assume the restoring torque is linear, but other functional forms are possible, e.g. $|\Omegac - \Omegas|^3$ when the superfluid is turbulent~\citep{GORTER1949285, Peralta2005}.

The stochastic torques $\xic(t)$ and $\xis(t)$ are treated as memoryless, white noise processes with
\begin{align}
\langle \xi_{\rm c,s}(t)\rangle &= 0\\
\label{eq:white_noise_torque_covariance}\langle \xi_{\rm c,s}(t)\xi_{\rm c,s}(t')\rangle &=\sigma^2_{\rm c,s}\delta(t-t'),
\end{align}
where $\langle\ldots\rangle$ denotes the ensemble average, and $\sigmac$ and $\sigmas$ are noise amplitudes.
Timing noise is often characterized in terms of the power spectral density of the phase residuals, $\delta \phi_{\rm c}(t)$, left over after subtracting the best-fit timing model from a set of times of arrival. The power spectral density is given by
\begin{align}
\Phi(f)
 &=
 \int_{-\infty}^\infty \textrm{d} \tau \, 
 e^{2\pi i f \tau}  
 \langle \delta\phi_{\rm c}(t) \delta\phi_{\rm c}(t+\tau) \rangle,
 \end{align}
which implies 
$\Phi(f)\propto f^{-4}$. We present an analytic calculation of the power spectrum of residuals for our model in detail in Appendix~\ref{app:power_spectral_analysis}, and compare that analytic calculation to a numerical simulation.
The timing noise model we discuss is a reasonable fit to the estimated timing noise for several accretion powered X-ray pulsars~\citep{10.1093/mnras/262.3.726,1993A&A...267..119B}, as well as many isolated radio pulsars (see, e.g. \citet{2019MNRAS.489.3810P,parthasarathy:2020wel,2020MNRAS.494..228L}, for recent timing noise studies). Such stochastic torques can arise physically due to, e.g. hydromagnetic instabilities in the accretion-disk-magnetosphere interaction in $\xic(t)$~\citep{romanova2004khj,romanova:2008ieu,dangelo:2010esf} or superfluid vortex avalanches in $\xis(t)$~\citep{Warszawski:2011ldo, drummond:2018esf}.

\subsection{State space representation}
\label{ssec:two_component_model:state_space_representation}
Equations (\ref{eq:crust_equation_of_motion})--(\ref{eq:white_noise_torque_covariance}) must be discretized in order to make contact with electromagnetic and gravitational-wave measurements of $\Omegac(t)$ and $\Omegas(t)$ sampled at discrete instances in time. The first step is to solve the linear, inhomogeneous, ordinary differential equations (\ref{eq:crust_equation_of_motion}) and (\ref{eq:superfluid_equation_of_motion}) analytically. Writing them in matrix form, one obtains
\begin{align}
\label{eq:ornstein_uhlenbeck}
    d\bm X &= \bm A\bm X \textrm{d}t + \bm N \textrm{d}t + \bm \Sigma \;\textrm{d}\bm B(t),
\end{align}
where $\bm X=(\Omegac,\Omegas)^T$ and $\bm N=(\Nc/\Ic, \Ns/\Is)^T$ are column vectors and we define
\begin{align}
    \bm A &= \begin{pmatrix}-1/\tauc & 1/\tauc\\1/\taus&-1/\taus\end{pmatrix}\\
    \bm \Sigma &= \begin{pmatrix}\sigmac/\Ic & 0\\0&\sigmas/\Is\end{pmatrix}.
\end{align}
The term $\textrm{d}\bm B(t)$ denotes a $2\times 1$ column vector containing increments of Brownian motion. Equation~(\ref{eq:ornstein_uhlenbeck}) is an Ornstein-Uhlenbeck process, and has a solution given by~\citet{gardiner_stochastic_methods}
\begin{align}
\label{eq:abstract_orn_uhl_solution}
    \bm X(t) = e^{\bm A t}\bm X(0) + \int_{0}^{t}e^{\bm A(t-t')}\bm N\,\textrm{d}t' + \int_0^{t}e^{\bm A(t - t')}\Sigma\,\textrm{d}\bm B(t'),
\end{align}
where $e^{\bm At}$ is a matrix exponential.

We now suppose we have some set of noisy measurements $\bm Y$ at times $t_1, \ldots t_{N_t}$, which we label $\bm Y_i = \bm Y(t_i).$ These measurements are related to the state variables $\bm X$ through the design matrix $\bm C$,
\begin{align}
\label{eq:measurement_equation}
\bm Y_i = \bm C \bm X(t_i) + \bm u_i,
\end{align}
where $\bm u_i$ represents the measurement noise sampled at the instant $t_i$. We take $\bm u_i$ to be zero mean with autocovariance $\langle \bm u_i \bm u_j^T\rangle = \delta_{ij}\bm R$ where $\delta_{ij}$ is the Kronecker delta. We use equation~(\ref{eq:abstract_orn_uhl_solution}) to construct a set of recursions that describe the state variables at time $t_{i+1}$ based on the state variables at time $t_i$:

\begin{align}
\bm X(t_{i+1}) &= \bm F_i \bm X(t_i) + \bm T_i + \bm \eta_i\label{eq:state_update_equation}\\
    \bm F_i &= e^{\bm A(t_{i+1} - t_i)}\label{eq:transition_integral}\\
    \bm T_i &= \int_{t_{i}}^{t_{i+1}} e^{\bm A(t_{i+1} - t')} \bm N \,\textrm{d}t'\label{eq:torque_integral}\\
    \bm \eta_i &= \int_{t_i}^{t_{i+1}} e^{\bm A(t_{i+1}-t')}\bm\Sigma\, \textrm{d}\bm B(t').\label{eq:noise_integral}
\end{align}
The exact analytic forms of~$\bm F_i$ and $\bm T_i$ are given by~\cite{exp_max_paper_meyers_melatos} for the case of uniform sampling, and are reproduced in Appendix~\ref{app:state_space_forms}. The covariance matrix of the noise term $\bm \eta_i$, given by
\begin{align}
\label{eq:process_noise_covariance}
\langle \bm \eta_i \bm \eta_j^T\rangle = \delta_{ij}\bm Q_i,
\end{align}
is also reproduced in Appendix~\ref{app:state_space_forms} (note that Einstein summation convention does not apply to the right-hand side of equation~(\ref{eq:process_noise_covariance})).

Given the linear measurement equation, (\ref{eq:measurement_equation}), the linear state update equation (\ref{eq:state_update_equation}), the covariance matrix of the process noise (\ref{eq:process_noise_covariance}), and the covariance matrix of the measurement noise $\bm R$, we can use a Kalman filter to track the state variables $\bm X$ through time.\footnote{By linear here we mean that the measurements depend linearly on the states, and the state at time $t_{i+1}$ depends linearly on the state at time $t_i$.}  We can also use the likelihood of the Kalman filter to perform Bayesian inference on the parameters $\tauc$, $\taus$, $\Nc/\Ic$, $\Ns/\Is$, $\sigmac/\Ic$, and $\sigmas/\Is$. The procedure for doing so is laid out in Section~\ref{sec:parameter_estimation}.

\cite{exp_max_paper_meyers_melatos} considered $\Delta t_i=t_i - t_{i-1} \ll \tauc, \taus, \Ic\Omegac/\Nc, \Is\Omegas/\Ns$, and made an Euler approximation $e^{\bm A\Delta t_i} \approx \mathbb I + \bm A\Delta t_i$ when calculating $\bm F$, $\bm T$ and $\bm Q$. In this paper we make no such approximation, leading to more complicated forms of $\bm F$, $\bm T$ and $\bm Q$. Importantly, the generalization allows us to analyze situations where the time between measurements is both non-uniform and the same, or longer than, the intrinsic time scales, $\tauc, \taus, \Ic\Omegac/\Nc, \Is\Omegas/\Ns$. These conditions are the norm in pulsar timing eperiments.

\subsection{Identifiability}
\label{ssec:two_component_model:identifiability}

We now turn to the question of whether, given a set of measurements $\{\bm Y_i\}_{i=1}^{i=N_t}$, we are able to estimate the unknown parameters of the two-component model described in equations (\ref{eq:crust_equation_of_motion})--(\ref{eq:white_noise_torque_covariance}). The mathematical structure of equations (\ref{eq:measurement_equation}) and (\ref{eq:state_update_equation}), which relate the measurements to the state and update the state respectively, can prevent certain model parameters from being estimated uniquely, no matter how plentiful and good the data. To check this, we must compare the number of independent conditions imposed on the data by the linear recursion relations in equations (\ref{eq:measurement_equation}) and (\ref{eq:state_update_equation}) (which grows with the number of data points) with the number of independent pieces of information in the data themselves (which also grows with the number of data points, albeit differently). 
This issue is known as identifiability in the statistical and engineering literature~\citep{BELLMAN1970329}. 

In the present application, data for $\Omegac$ can be obtained from radio or X-ray timing~\citep{lyne_graham-smith_2012}. Data for $\Omegas$ are harder to obtain, as they rely on directly measuring the rotational state of the interior, which is decoupled from electromagnetic observables. However, if the superfluid has a time-varying quadrupole moment, and therefore emits gravitational waves~\citep{alpar:1996aew,sedrakian:2002vbf,jones:2006aok,2019AIPC.2127b0007H,melatos:2010bvy,Melatos2014}, it is straightforward to relate $\Omegas$ to measurements of the gravitational wave emission frequency, e.g. from hidden Markov model tracking~\citep{2016PhRvD..93l3009S,2017PhRvD..96j2006S,2018PhRvD..97d3013S}. Continuous gravitational radiation has not been detected yet from a rotating neutron star~\citep{2013PrPNP..68....1R}, but there is every hope this will change soon.
For the rest of this paper, we consider two observational scenarios: (1) a future scenario, in which we have gravitational-wave and electromagnetic measurements that independently probe $\Omegas$ and $\Omegac$ respectively, meaning that $\bm C$ in equation~(\ref{eq:measurement_equation}) is the identity matrix; and (2) an existing scenario, in which we have only electromagnetic measurements of $\Omegac$, meaning that $\bm Y_i$ is a scalar at each time-step $i$, and $\bm C = (1, 0)$ is a row-vector. It is straightforward to extend these cases to parameterize $\bm C$ as well, but that falls outside the scope of this paper.

In the rest of this section, we analytically calculate which parameters out of $\tauc$, $\taus$, $\Nc/\Ic$ and $\Ns/\Is$ we expect to measure in the two scenarios discussed above.
We consider the problem with no measurement or process noise, because it is simpler to deal with and can yield important analytical insights. If a problem is tractable in this context, then it should also be tractable in the situation where white noise is present, as long as there are enough data.

\subsubsection{Electromagnetic and gravitational-wave observations}
\label{sssec:identifiability:em_gw_observations}

In the future scenario we assume we measure both $\Omegac(t_i)$ and $\Omegas(t_i)$ directly at $t_1, \ldots t_{N_t}$. Assuming that there is no process or measurement noise, equations (\ref{eq:crust_equation_of_motion}) and (\ref{eq:superfluid_equation_of_motion}) reduce to
\begin{align}
    \frac{\textrm{d}\Omegac}{\textrm{d}t} &= \frac{\Nc}{\Ic} - \frac{1}{\tauc}\left(\Omegac - \Omegas\right)\label{eq:crust_equation_of_motion_nonoise},\\
    \frac{\textrm{d}\Omegas}{\textrm{d}t} &= \frac{\Ns}{\Is} - \frac{1}{\taus}\left(\Omegas - \Omegac\right)\label{eq:superfluid_equation_of_motion_nonoise},
\end{align}
and equation (\ref{eq:measurement_equation}), reduces to 
\begin{align}
\bm Y_i = \bm C \bm X(t_i).
\end{align}
In this simplified analysis, when a state is measured it is assumed to be exactly known. In practice the measurements are known but the true states are unknown because there is measurement error.

If we have $N_t$ measurements of $\Omegac$ and $\Omegas$, then the recursion relations defined in~equation (\ref{eq:state_update_equation}) yield $2N_t - 2$ equations with four unknowns, $\tauc$, $\taus$, $\Nc/\Ic$ and $\Ns/\Is$. We can solve for all four unknowns as long as we have $N_t \geq 3$ and the equations are all independent.

It is easier to analyse a continuous version of this system. Instead of having a discrete list of $N_t$ data points we assume there is a continuous function $\Omegac(t)$ and its derivatives $\Omegac^{(1)}$ to $\Omegac^{(N_t-1)}$ are known at a particular point. These are equivalent problems because knowing $\Omegac$ at two points allows $\dot \Omega_{\rm c}$ to be estimated, knowing it at three points allows $\ddot \Omega_{\rm c}$ to be estimated, and so on.
Rearranging equations (\ref{eq:crust_equation_of_motion_nonoise}) and (\ref{eq:superfluid_equation_of_motion_nonoise}) yields
\begin{align} 
\begin{bmatrix}
\dot \Omega_{\rm c} \label{eq:em_gw_identifiability} \\
\dot \Omega_{\rm s} \\
\ddot \Omega_{\rm c} \\
\ddot \Omega_{\rm s}
\end{bmatrix}
=
\begin{bmatrix} 
\Omega_{\rm s} - \Omega_{\rm c} & 0 & 1 & 0\\
0 & \Omega_{\rm c} - \Omega_{\rm s} & 0 & 1\\
\dot \Omega_{\rm s} - \dot \Omega_{\rm c} & 0 & 0 & 0\\
0 & \dot \Omega_{\rm c} - \dot \Omega_{\rm s} & 0 & 0
\end{bmatrix}
\begin{bmatrix}
1/\tauc \\
1/\taus \\
\Nc/\Ic \\
\Ns/\Is
\end{bmatrix}.
\end{align}
A maximum-likelihood estimate of the unknown parameters in the column vector on the right-hand side of equation~(\ref{eq:em_gw_identifiability}) can be found in terms of the data on the left-hand side of equation~(\ref{eq:em_gw_identifiability}) as long as the matrix on the right-hand side of equation~(\ref{eq:em_gw_identifiability}) is invertible. The condition for invertibility is $\dot\Omega_{\rm c}\neq\dot\Omega_{\rm s}$. In practice, $\xi_{\rm c}(t)$ and $\xi_{\rm s}(t)$ are constantly inducing fluctuations and preventing the two components reaching equilibrium. As a result, the condition for invertibility will always be satisfied except at discrete instants, when one has $\dot\Omega_{\rm c}=\dot\Omega_{\rm s}$ instantaneously by chance. Therefore, given measurements of both $\Omegac(t)$ and $\Omegas(t)$ we should be able to estimate the parameters $\tauc$, $\taus$, $\Nc/\Ic$ and $\Ns/\Is$.
\subsubsection{Electromagnetic only observations}
\label{sssec:identifiability:em_only_observations}
Until some timing signature that tracks the angular velocity of the superfluid is available (e.g. continuous gravitational waves), we are obliged to rely on electromagnetic data tied to the crust, viz. $\Omegac(t_1)\ldots\Omegac(t_{N_t})$.
There are now $N_t+4$ unknowns -- the four parameters mentioned above, as well as $\Omegas(t_1) ,\ldots,\Omegas(t_{N_t})$. There are still $2N_{t}-2$ equations in the discrete picture, meaning we need $N_t \geq 6$ in order to close the system. However, the $2N_{t}-2$ equations do not all yield independent information. Moreover, only having access to the trajectory of $\Omegac(t)$ means that we are sensitive only to certain combinations of parameters. This is most easily seen in the continuous picture, where we can do something similar to equation (\ref{eq:em_gw_identifiability}). If we take a time-derivative of equation~(\ref{eq:crust_equation_of_motion_nonoise}) and use equation~(\ref{eq:superfluid_equation_of_motion_nonoise}) to substitute for $\dot\Omega_{\rm s}$, it is straightforward to find
\begin{align}
\label{eq:em_only_omgddot}
\ddot\Omega_{\rm c} &= -\left(\frac{1}{\tauc} + \frac{1}{\taus} \right)\dot\Omega_{\rm c} + \frac{\Nc}{\taus\Ic} + \frac{\Ns}{\tauc\Is}.
\end{align}
This equation determines the evolution of $\Omegac(t)$ without any reference to $\Omegas$. So given only measurements of $\Omegac$, only the combinations of parameters that appear in (\ref{eq:em_only_omgddot}), namely
\begin{align}
\label{eq:reduced_relaxation_time}
\tau &= \frac{\tauc\taus}{\tauc+\taus}
\end{align}
and
\begin{align}
\label{eq:ensemble_averaged_spindown}\langle \dot\Omega_{\rm c}\rangle &= \frac{1}{\tauc+\taus}\left(\tau_{\rm c} \frac{\Nc}{\Ic}+\taus \frac{\Ns}{\Is}\right),
\end{align} can be determined. 
This indicates that the $2N_t-2$ equations are not independent.

As discussed by~\cite{exp_max_paper_meyers_melatos}, $\tau$ and $\langle \dot\Omega_{\rm c} \rangle$ are the reduced relaxation time-scale and the long-term, ensemble-averaged, secular spin-down of the system.
Sure enough,~\cite{exp_max_paper_meyers_melatos} found that $\tau$ and $\langle\dot\Omega_{\rm c}\rangle$ are estimated well from the time series, $\Omegac(t_1)\ldots \Omegac(t_{N_t})$, in Monte Carlo trials, even though $\Nc$, $\Ns$, $\tauc$ and $\taus$ cannot be estimated accurately on an individual basis.

\section{Parameter estimation}
\label{sec:parameter_estimation}
In this section we present a method for Bayesian estimation of the unknown parameters, $\bm\theta=(\tauc,\taus,\Nc/\Ic,\Ns/\Is, \sigmac/\Ic,\sigmas/\Is)$, using a likelihood that can be evaluated quickly and reliably with a Kalman filter. 

The problem of estimating parameters of a linear dynamic system has a wide variety of solutions. Many of those solutions, motivated by real-time engineering applications, are optimized for speed. Accuracy is pursued only insofar as it improves performance of a control system or tracking of a set of state-variables through time. Indeed, the maximum-likelihood method presented in~\cite{exp_max_paper_meyers_melatos} returns parameter estimates in $\mathcal O(\textrm{seconds})$. However, it does not return posterior distributions or an obvious method for characterizing uncertainties on the parameters. 

In this paper, tracking and control are secondary and parameter estimation is the goal. We seek to do precision modelling of neutron stars. Motivated by this, we focus on a method that requires many likelihood calculations, and therefore takes $\mathcal O(\textrm{minutes})$ -- $\mathcal O(\textrm{hours})$ to complete on, e.g. a 2.3 GHz dual-core processor, but returns full posterior probability distributions of the unknown parameters. In Section~\ref{ssec:kalman_and_likelihood} we discuss the Kalman filter and its associated likelihood function. In Section~\ref{ssec:mcmc_methods} we discuss how we can use MCMC methods to estimate the posterior distribution of the parameters, $\bm \theta.$

\subsection{Kalman filter and its likelihood}
\label{ssec:kalman_and_likelihood}
The Kalman filter \citep{kalman1960} is a recursive algorithm used to estimate a set of unknown state-variables, $\bm X$ based on a set of noisy measurements, $\bm Y$. The traditional Kalman filter assumes Gaussian disturbances (known as process noise) on $\bm X$, Gaussian measurement error on $\bm Y$,  a linear relationship between $\bm Y$ and $\bm X$ and a linear recursion for updating $\bm X$. In short, the assumptions made by the Kalman filter are the assumptions underpinning equations (\ref{eq:measurement_equation}) and (\ref{eq:state_update_equation}), making it an ideal tool. Extensions to non-linear state-transitions [i.e. $\bm F\bm X \rightarrow f(\bm X)$ in equation (\ref{eq:state_update_equation})], and non-linear measurement equations [i.e. $\bm C\bm X \rightarrow c(\bm X)$ in equation (\ref{eq:measurement_equation})] can be achieved using a range of tools such as the extended Kalman filter~\citep{jazwinski1970}, the unscented Kalman filter~\citep{julier1997,Wan00theunscented,julier2004}, or particle filters~\citep{DELMORAL1997653}.

The full set of Kalman recursions is shown in Appendix~\ref{app:kalman_recursions}. The output of the filter is a set of estimates of the state variables, $\hat{\bm X}_i$, and the covariance matrix of those estimates, $\bm P_{i}$, for each time-step, $i=1\ldots N_t$. Through equation (\ref{eq:measurement_equation}) it is straightforward to produce an expectation of the measurements, $\hat{\bm Y}_i$, at each time-step (as discussed in Appendix~\ref{app:kalman_recursions}, they are calculated as part of the filter recursions). The error in the measurement estimate, $\bm \epsilon_i = \bm Y_i - \hat{\bm Y}_i$, is known as the ``innovation.'' The innovation has an associated covariance matrix, $\langle \bm \epsilon_i\bm \epsilon_i^T\rangle = \bm S_i$, that is used to calculate the Kalman filter likelihood
\begin{align}
\label{eq:kalman_likelihood}
\log p(\{\bm Y_i\}_{i=1}^{\bm N_t} | \bm \theta) = -\frac{1}{2}\sum_{i=1}^{N_t}\left[N_d\log(2\pi) + \log|\bm S_i| + \bm \epsilon_i^T \bm S_i^{-1}\bm \epsilon_i\right],
\end{align}
where $N_d$ is the dimension of $\bm Y_i$ and the Einstein summation convention does not apply to the right-hand side of equation~(\ref{eq:kalman_likelihood}). We have $N_d=2$ when we have electromagnetic and gravitational-wave measurements or $N_d=1$ when we have only electromagnetic measurements. The dependence of equation~(\ref{eq:kalman_likelihood}) on $\bm\theta$ indicates that we make a specific choice of $\bm\theta$ when running the Kalman filter and calculating the log-likelihood. A discussion of the Kalman filter likelihood is given in Appendix~\ref{app:kf_log_likelihood}.

\subsection{Posterior distributions}
\label{ssec:mcmc_methods}
We can combine the likelihood in equation~(\ref{eq:kalman_likelihood}) with a prior distribution on the parameters, $p(\bm \theta)$, to estimate the posterior on $\bm\theta$ using Bayes' Rule
\begin{align}
\label{eq:bayes_rule}
    p(\bm\theta|\{\bm Y_i\}) = \frac{p(\{\bm Y_i\} | \bm \theta)p(\bm\theta)}{p(\{\bm Y_i\})},
\end{align}
where we suppress the range of indices $i=1\ldots N_t$ for brevity. The denominator represents the Bayesian evidence and is found by marginalizing over the likelihood, weighted by the prior:
\begin{align}
p(\{\bm Y_i\}) = \int \textrm{d}\bm\theta\,p(\{\bm Y_i\} | \bm \theta)p(\bm\theta).
\end{align}

There are six parameters of interest, and so a brute-force evaluation of equation~(\ref{eq:bayes_rule}) might be feasible, if the mode of the distribution can be found first using the methods presented in~\citet{exp_max_paper_meyers_melatos}. However, a gridded approach with 100 points in each parameter requires $\sim 10^{12}$ likelihood evaluations, which is computationally intensive, and may suffer biases in some circumstances, e.g. posteriors with multiple modes. The computational cost is tolerable for a single pulsar in principle but grows prohibitive when tracking many pulsars on a regular basis.
 Therefore, we use a nested sampling approach~\citep{skilling2006} to estimate the posterior distribution and the Bayesian evidence. While we do not interpret the evidence in this paper, it can be useful for performing model selection. We use the \texttt{dynesty}~\citep{speagle2020oas} nested sampler through the \texttt{Bilby} front-end~\citep{Ashton:2018jfp}. We discuss prior distributions in the next sub-sections.

\begin{figure*}
    \centering
    \includegraphics[width=0.49\textwidth]{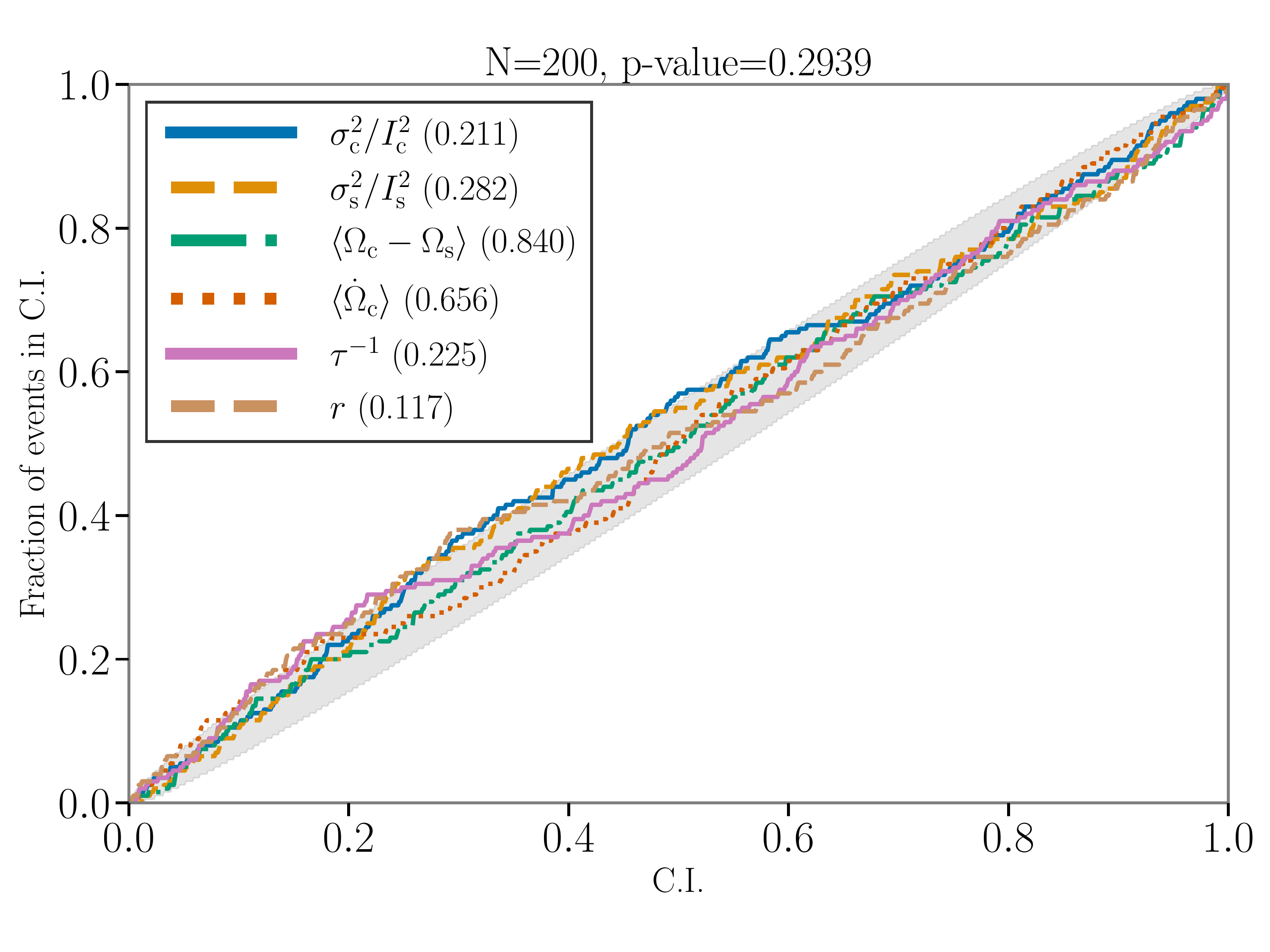}
    \includegraphics[width=0.49\textwidth]{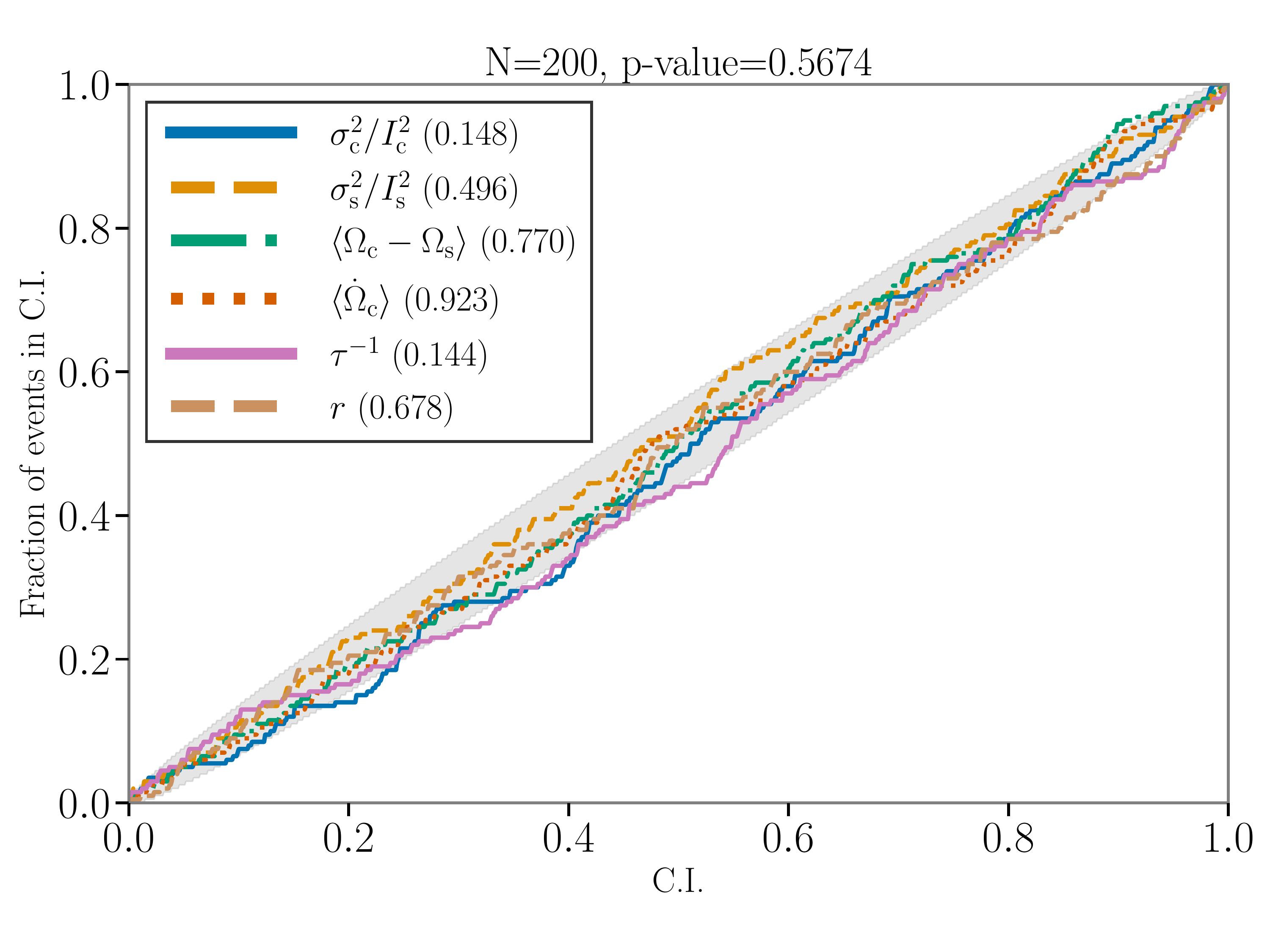}
    
    \caption{PP plots for present-day observational scenario. Each colored curve corresponds to one parameter (color code in the legend). It indicates the fraction of 200 simulated events events whose injected parameter value falls within a certain confidence interval (label C.I.) as a function of that confidence interval. If the posterior distributions are properly estimated, then the curves should fall within the shaded region. The $p$-values in the legend indicate whether the posteriors associated with each parameter are well-behaved (a more technical discussion is given in the text). {\bf Left:} Present-day scenario with electromagnetic measurements only and $N_t=600$ spread over 1825 days; {\bf Right:} same as for bottom left but with $N_t=1200$ spread over 3650 days.}
    \label{fig:pp_plot_em}
\end{figure*}

T he results presented below are insensitive to the choice of sampler settings. For example, one of the main tunable features in nested sampling is the number of `live points.' In nested sampling, the live point with the lowest likelihood value is replaced by a new point in each step of the sampler. For large or multi-modal parameter spaces, increasing the number of live points can greatly improve sampler performance by guaranteeing that the full parameter space is explored. We produce nearly identical posterior results using 200 live points, 500 live points, and 1000 live points in the nested sampling for the parameter choices in Table~\ref{tab:parameters_and_priors}. The results for probability-probability (PP) plots presented in Figures~\ref{fig:pp_plot_em_gw} and~\ref{fig:pp_plot_em} produce reasonable results for both 200 and 500 live points. We use 500 live points for all of the results presented below.

\subsection{Parameter combinations}
\label{ssec:choosing_parameters_and_priors}
The choice of parameters, $\bm \theta$, and their prior distribution, $p(\bm\theta)$, depend on the situation.
Instead of sampling $\bm\theta=(\tauc,\taus,\Nc/\Ic,\Ns/\Is,\sigmac/\Ic,\sigmas/\Is)$, we can  transform to a more appropriate set of parameters that have a natural physical interpretation and fewer degeneracies between them. For example, the product $\tauc\Nc/\Ic$ in equation~(\ref{eq:ensemble_averaged_spindown}) leads to a degeneracy in $\tauc$ and $\Nc/\Ic$, which is evident in the results in~\cite{exp_max_paper_meyers_melatos}. Therefore we favor parameters that show up in equations (\ref{eq:transition_matrix_full})--(\ref{eq:process_noise_covariance_full}), which have the added benefit of having simple physical interpretations. We continue to use $\sigmac^2/\Ic^2$ and $\sigmas^2/\Is^2$, as well as 
\begin{align}
   \tau &= \frac{\tauc\taus}{\tauc+\taus} \\
   r &= \frac{\taus}{\tauc}\\
   \langle\dot\Omega_{\rm c}\rangle &= \frac{1}{\tauc + \taus}\left(\tauc\frac{\Nc}{\Ic} + \taus\frac{\Ns}{\Is}\right)\\
   \langle\Omegac - \Omegas\rangle &= \tau\left(\frac{\Nc}{\Ic} - \frac{\Ns}{\Is}\right).
\end{align}
In this representation, $\tau$ is the relaxation time of the system after a perturbation in the lag, $\Omegac - \Omegas$; $r$ is the ratio of the two relaxation times, which equals $\Is/\Ic$ when the final terms of equations (\ref{eq:crust_equation_of_motion}) and (\ref{eq:superfluid_equation_of_motion}) form an action-reaction pair; $\langle\dot\Omega_{\rm c}\rangle$ is the ensemble-averaged spin-down of the system; and $\langle\Omega_{\rm c} - \Omega_{\rm s}\rangle$ is the ensemble-averaged steady-state lag. It is straightforward to move between this new set of parameters and the original set of parameters.

Sometimes we might wish to set a physical prior probability distribution, $p(\bm\theta)$, on some set of parameters, $\bm\theta$, but wish to estimate a different set of parameters, $\bm\theta' = g(\bm\theta)$. In this scenario, $p(\bm\theta')$ is estimated using the Jacobian of the transformation $g(\cdot)$,
\begin{align}
p(\bm\theta') &= p[g^{-1}(\bm\theta')]/|\det \bm J|,\\
J_{ij}&=\frac{\partial \theta'_i}{\partial \theta_j}.
\end{align}

To illustrate how one might choose a set of parameters, $\bm\theta$,  we present two practical scenarios one might encounter in pulsar astronomy. There are, of course, more permutations, but the logic extends to those situations as well.

\subsubsection{Case I: isolated, non-accreting radio pulsar}
A non-accreting radio pulsar is believed to have $\Nc < 0$ and $\Ns<0$~\citep{2015IJMPD..2430008H}. Moreover, radio timing data indicate $\langle\dot\Omega_{\rm c}\rangle < 0$, except during a glitch. Nearly all studies of the two-component model and standard glitch models indicate $\langle \Omegac - \Omegas\rangle<0$ in this situation as well~\citep{2015IJMPD..2430008H}. For example, in standard glitch models, the crust spins up in response to an impulsive deceleration of the core. Therefore, in this situation we would choose $\bm\theta = (\tau^{-1}, r, \langle\dot\Omega_{\rm c}\rangle, \langle\Omegac - \Omegas\rangle, \sigmac^2/\Ic^2, \sigmas^2/\Is^2)$ and set appropriate priors to restrict $\langle\dot\Omega_{\rm c}\rangle < 0$, and $\langle\Omegac - \Omegas\rangle \leq 0$. 

We choose $\tau^{-1}$ instead of $\tau$ because it is $\tau^{-1}$ that appears more readily in the transition matrix and the exponential decay of perturbations in $\Omegac-\Omegas$. Meanwhile, we sample over the square of the stochastic torque amplitudes because these are what show up naturally in the process noise covariance matrix $\bm Q$, shown in (\ref{eq:process_noise_covariance_full}).

\subsubsection{Case II: accreting X-ray pulsars, visibly spinning-down}
As discussed in Section~\ref{sec:two_component_model}, when a system is accreting and spinning down, one may have $\Nc < 0$ or $\Nc > 0$, and $\langle\Omegac-\Omegas\rangle$ can take either sign. However if X-ray timing data imply $\langle\dot\Omega_{\rm c}\rangle < 0$, and if we make the physically motivated assumption $\Ns<0$, then it stands to reason that a good parameter set is $\bm\theta = (\tau^{-1}, r, \langle\dot\Omega_{\rm c}\rangle,\Ns/\Is, \sigmac^2/\Ic^2, \sigmas^2/\Is^2)$, on which we can set reasonable physical priors.

\subsection{Prior distributions}
Once we choose a set of parameters, $\bm\theta$, it remains to choose sensible prior distributions, $p(\bm\theta)$, on those parameters. In the rest of this section we go through each of the physical parameters discussed previously and address the range they could possibly take based on astrophysical observations and/or theoretical arguments. In the next section, we test our model across that parameter space. The final distributions are presented in Table~\ref{tab:parameters_and_priors}.

If a maximum-likelihood fit exists for $\langle \dot\Omega_{\rm c}\rangle$ using the same data on which we plan to do our tracking, e.g. supplied by the algorithm in~\cite{exp_max_paper_meyers_melatos}, then a sensible prior would be a uniform prior that allows for ample excursion from the maximum-likelihood value. A Gaussian prior centered on the maximum-likelihood value with uncertainty given by the error in the maximum-likelihood fit is not permissible, as this would be using the data twice and would artificially narrow the posterior distribution. Throughout the rest of this paper we consider pulsars with spin-downs in the range $-10^{-10}~\mathrm{rad~s^{-2}}\leq\langle \dot\Omega_{\rm c}\rangle< 0$, as this encompasses a broad range of millisecond and young pulsars.

For the relaxation time, $\tau$, in some cases an estimate might already exist, e.g. from the post-glitch recovery of the spin period. In the absence of such a measurement, one can choose a prior distribution informed by the population of glitch relaxation time measurements that have already been made. In~\cite{yu:2016oai}, the authors report 89 glitch relaxation times ranging from 0.5 days to 1000 days. Based on this, we use a log-uniform prior on $\tau^{-1}$ between $10^{-8}~\rm{s^{-1}}$ and $10^{-5}~\rm{s^{-1}}$ throughout the rest of this paper.

The ratio of timescales, $r$, is equivalent to $\Is/\Ic$ when~(\ref{eq:crust_equation_of_motion}) and (\ref{eq:superfluid_equation_of_motion}) form an action-reaction pair. The  literature generally considers two cases for $\Ic/\Is$, both of which are typically framed in the context of glitches. In the first, the crustal lattice is locked in corotation with the neutrons in the core and the proton-electron fluid via vortex-fluxoid interactions. In this case, the angular momentum transferred during a glitch is typically stored in the inner-crust superfluid. Under this model, one typically takes $\Ic/\Is \sim 10^{2}$~\citep{link:1999iek,lyne:2000cie,2011MNRAS.414.1679E}.  In the second scenario, the angular momentum transferred during the glitch is stored in the superfluid components of the core itself~\citep{chamel:2012cwl}. In this case, one has $\Ic/\Is \ll 1$. Throughout the rest of this paper, we consider a log-uniform prior on $r$ ranging from $10^{-2}$ to $10^2$. This gives equivalent weight to all possible values in between, as opposed to favoring $\Ic/\Is > 1$, as a standard uniform prior would.

In~\cite{exp_max_paper_meyers_melatos}, a comparison is made between the noise parameters $\sigmac$ and $\sigmas$ and one of the standard timing noise statistics in the literature~\citep{1980ApJ...237..216C}. Timing noise varies across the pulsar population and depends on whether the pulsar is young, whether it is spinning down quickly, or whether it is accreting. In this paper we choose log-uniform priors on $\sigmac^2/\Ic^2$ and $\sigmas^2/\Is^2$ in the range $10^{-24}-10^{-16}~\rm{rad^2~s^{-3}}$, which is consistent with noisy, young pulsars and magnetars~\citep{2019MNRAS.485....2C,exp_max_paper_meyers_melatos}. Moving to pulsars where the timing noise is smaller, or there is no confident estimate of a red-noise process, would require a model selection framework that we do not develop here.

Finally, we consider the ensemble-averaged lag $\langle \Omegac(t) - \Omegas(t)\rangle$. For $\Nc/\Ic \approx \Ns/\Is$, the lag tends to zero. To place an upper limit on $|\langle \Omegac(t) - \Omegas(t)\rangle|$, first assume $|\langle \Omegac(t) - \Omegas(t)\rangle| \ll \Omega_{\rm c}(t)$ and look at the two limiting cases, $\Nc/\Ic \gg \Ns/\Is$ and $\Nc/\Ic \ll \Ns/\Is$. In the first case, $\Nc/\Ic \gg \Ns/\Is$, we find
\begin{align}
\label{eq:lag_nc_gg_ns}
|\langle \Omegac(t) - \Omegas(t)\rangle| \approx \frac{\tau}{2\tau_{A}} \left(1+r\right)\Omegac,
\end{align}
where $\tau_{A} = \Omegac / 2\dot\Omega_{\rm c}$ is the characteristic age of the pulsar. In the second case $\Nc/\Ic \ll \Ns/\Is$, we find
\begin{align}
\label{eq:lag_nc_ll_ns}
|\langle \Omegac(t) - \Omegas(t)\rangle| \approx \frac{\tau}{2\tau_{A}} \left(\frac{1+r}{r}\right)\Omegac.
\end{align}
Applying the constraints on $r$ discussed above, we find that the right hand sides of (\ref{eq:lag_nc_gg_ns}) and (\ref{eq:lag_nc_ll_ns}) generally fall within the range of $10^{-5}\Omegac-10^{-3}\Omegac$ for the young glitching pulsars considered in~\cite{yu:2016oai}. Therefore, for the rest of this paper, we take
\begin{align}
|\langle \Omegac(t) - \Omegas(t)\rangle|\leq 10^{-3}\Omegac(t). 
\end{align}
This lag is consistent with what is accommodated by some continuous gravitational-wave searches~\citep{abbott:2008cnw,abbott:2019pli}. As discussed in Section~\ref{ssec:choosing_parameters_and_priors}, for $\Nc < 0$ and $\Ns< 0$ (as for an isolated pulsar), the lag is negative.

\section{Validation with synthetic data}
\label{sec:results}

\subsection{Synthetic data generation}
\label{ssec:synthetic_data_generation}
We generate synthetic time-series, $\Omegac(t_1)\ldots\Omegac(N_{t})$ and $\Omegas(t_1)\ldots\Omega_s(N_{t})$ by integrating the equations of motion (\ref{eq:crust_equation_of_motion}) and (\ref{eq:superfluid_equation_of_motion}) numerically. To do the integration we use the Runge-Kutta It\^o integrator~\citep{10.2307/41062628} in the \texttt{sdeint} python package\footnote{\url{https://github.com/mattja/sdeint}}. We consider data sets that are either 5 years or 10 years in length with $N_{t}=600$ and $N_{t}=1200$ measurements respectively. The measurement times are not uniformly sampled over the 5 or 10 year period in order to mimic a realistic observational campaign. Instead we draw $N_t$ observations randomly from a time-series sampled hourly. The observing cadence we choose above is reasonable for campaigns using the UTMOST instrument on the Molonglo Observatory Synthesis Telescope~\citep{2017PASA...34...45B} and the Canadian Hydrogen Intensity Mapping Experiment (CHIME)~\citep{10.1117/12.2054950}, both of which can observe pulsars daily as they transit across the sky. In Appendix~\ref{app:less_frequent_observations} we also consider $N_t=600$ measurements spaced over a 20 year period, which is more representative of the steerable telescopes often used for pulsar timing. We set the Gaussian measurement error at the level of $\bm R = \mathbb I \times 10^{-18}~\rm{rad^2~s^{-2}}$ for all simulated measurements (where $\mathbb I$ is the identity matrix of dimension $N_d\times N_d$). Throughout all of the simulations we fix $\Omegac(t_1) = 10~\rm{rad~s^{-1}}$, and $\Omegas(t_1) = \Omegac(t_1) - \langle\Omegac - \Omegas\rangle$. 

\subsection{Representative example}
\label{ssec:results_on_a_simple_example}
We begin the characterization of this method with a simple example: an isolated neutron star that is spinning down. The parameters used for this injection are shown in the ``Injected Value'' column of Table~\ref{tab:parameters_and_priors}. The first six rows show the underlying model parameters, while the last four indicate the derived quantities we seek to infer with the MCMC. As discussed in Section~\ref{ssec:choosing_parameters_and_priors}, for an isolated neutron star we sample over $\bm\theta=(\tau^{-1}, \taus/\tauc, \langle\dot\Omega_{\rm c}\rangle, \langle \Omegac - \Omegas\rangle, \sigmac^2/\Ic^2, \sigmas^2/\Is^2).$ The prior probability distributions used for the sampling are given in the fourth column of Table~\ref{tab:parameters_and_priors}.

\begin{figure*}
    \centering
    \includegraphics[width=0.7\textwidth]{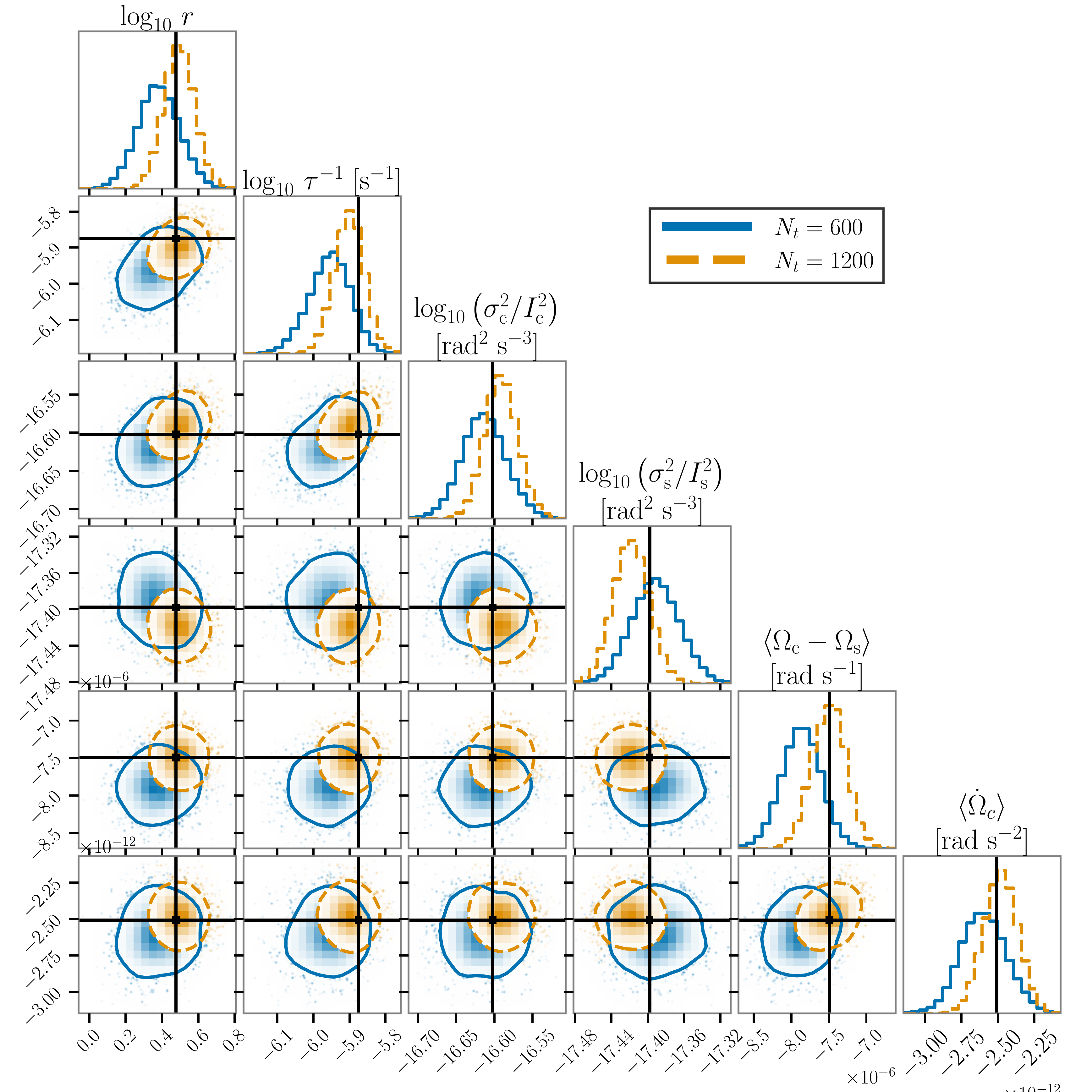}
    \caption{Posterior distributions for the future electromagnetic and gravitational-wave measurement scenario. Blue, solid lines and contours indicate $N_t=600$ and orange, dashed lines indicate $N_t=1200$. Solid, black, horizontal and vertical lines indicate the true injected values, which can be found in Table~\ref{tab:parameters_and_priors}. The contours indicate 90\% confidence levels. We accurately estimate each of the parameters of interest, and when we include more data the posterior distributions tighten.}
    \label{fig:posteriors_em_gw_simple_example}
\end{figure*}

First, we consider the future scenario where we have both electromagnetic and gravitational-wave measurements. We show the full posterior distributions for each of the parameters in Fig.~\ref{fig:posteriors_em_gw_simple_example}. The blue, solid posteriors are for $N_t=600$ spread over 5 years, while the orange, dotted ones are for $N_t=1200$ spread over 10 years. The thick vertical or horizontal black lines indicate the injected values. In the 2D posterior plots, the contours indicate 90\% confidence levels. It is clear that for this example, we are able to accurately estimate the parameters. As we add more data, the posterior distributions narrow; this can be seen by comparing the blue solid posteriors, $N_t=600$, to the orange dashed posteriors, $N_t=1200$.

\begin{figure*}
    \centering
    \includegraphics[width=0.7\textwidth]{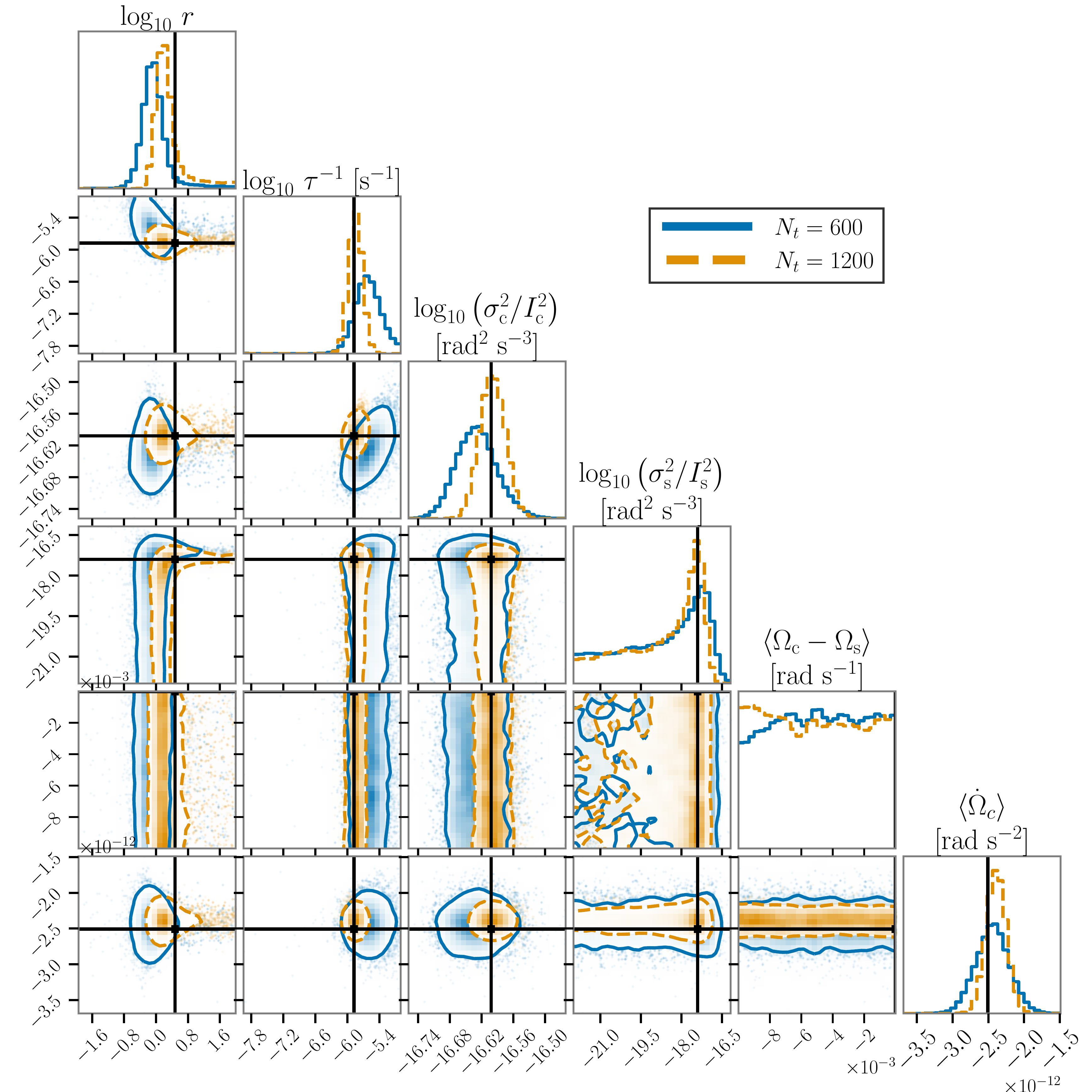}
    \caption{Posterior distributions for the present-day electromagnetic-only measurement scenario. Blue, solid lines and contours indicate $N_t=600$ and orange, dashed lines indicate $N_t=1200$. Solid, black, horizontal and vertical lines indicate the true injected values, which can be found in Table~\ref{tab:parameters_and_priors}. The contours indicate 90\% confidence levels. We see that for the $N_t=600$ data set the only parameters accurately estimated are $\sigmac^2/\Ic^2$ and $\langle \dot\Omega_{\rm c}\rangle$. For the $N_t=1200$ data set the posterior has a peak near the true value of $\tau^{-1}$, as well as $\sigmas^2/\Is^2$.}
    \label{fig:posteriors_em_only_simple_example}
\end{figure*}

Next, we consider the present-day scenario where we have only electromagnetic measurements. We show full posterior distributions for each parameter in Fig.~\ref{fig:posteriors_em_only_simple_example}. The color and line-style conventions are the same as for the left-hand panel. In this case, for $N_t=600$ the only parameters that are accurately estimated are $\sigmac^2/\Ic^2$,  $\langle\dot\Omega_{\rm c}\rangle$, and $\tau^{-1}$. Meanwhile, for $N_t=1200$, a peak starts to form near the true value of $\sigmas^2/\Is^2$. There is no evidence in Fig.~\ref{fig:posteriors_em_only_simple_example} that we are able to constrain $\langle \Omegac - \Omegas\rangle$ at all --- something that is consistent with the identifiability discussion in Section~\ref{ssec:two_component_model:identifiability}.

In both the future and current scenarios, we are able to constrain $r$ to within an order of magnitude, which is unexpected given the identifiability analysis in Section~\ref{sssec:identifiability:em_only_observations}. It is possible that a more sophisticated identifiability analysis would find that this is plausible. For example, an analysis that includes stochastic torques might find, e.g. a term related to $\xi_s$ (or its derivatives) in (\ref{eq:em_only_omgddot}), which could offer insight into $\sigmas/\Is$ and $r$. 

\bgroup
\def\arraystretch{1.5}
\begin{table*}
    \centering
\begin{tabular}{c|l|r|l|l}
\hline
    Parameter & Units & Injected Value (Section~\ref{ssec:results_on_a_simple_example}) & Prior (Section~\ref{ssec:results_on_a_simple_example}) & Prior (Section~\ref{ssec:results_full_parameter_space})\\
    \hline\hline
    $\tauc$ &
    $\rm{s}$ &$10^{6}$ & 
    &
    \\
    $\taus$
    & $\rm{s}$
    & $3\times 10^{6}$&
    \\
    $\frac{\Nc}{\Ic}$ &
    $\rm{rad~s^{-2}}$ &$-10^{-11}$&
    &
    \\
    
    $\frac{\Ns}{\Is}$ &
    $\rm{rad~s^{-2}}$ &$-10^{-14}$&
    &
    \\
    
$\frac{\sigmac^2}{\Ic^2}$ &
$\rm{rad^2~s^{-3}}$ &$2.5\times10^{-17}$&
$\log\mathcal{U}(10^{-24}, 10^{-16})$ &
$\log\mathcal{U}(10^{-24}, 10^{-16})$
\\

$\frac{\sigmas^2}{\Is^2}$  &
$\rm{rad^2~s^{-3}}$ &$4\times10^{-18}$&
$\log\mathcal{U}(10^{-24}, 10^{-16})$ & 
$\log\mathcal{U}(10^{-24}, 10^{-16})$\\
    \hline
    $r$ & &
    3 &
    $ \log \mathcal U(10^{-2}, 10^2)$ &
    $ \log \mathcal U(10^{-2}, 10^2)$\\
    
    $\tau^{-1}$ &
    $\rm{s^{-1}}$ & $1.3 \times 10^{-6}$&
    $\mathcal U(10^{-8}, 10^{-5})$ &
    $\mathcal U(10^{-8}, 10^{-5})$\\
    
    $\langle \dot\Omega_{\rm c,s}\rangle$ & 
    $\rm{rad~s^{-2}}$ &$-2.51\times10^{-12}$&
    $\mathcal{U}(-10^{-10}, 0)$ &
    $\mathcal{U}(-10^{-10}, 0)$\\
    
    $\langle \Omega_{\rm c} - \Omega_{\rm s}\rangle$ &
    $\rm{rad~s^{-1}}$ &$-7.5\times 10^{-6}$&
    $\mathcal{U}(-10^{-2}, 0)$ &
    $\mathcal{U}(-10^{-2}, 10^{-2})$\\
    \hline
    \end{tabular}
    \caption{Parameters used in the tests with synthetic data in Section~\ref{sec:results}. Above the horizontal line are the underlying parameters that define the model. Below the horizontal line are derived parameters inferred by the MCMC algorithm, whose physical meaning are discussed in the text. The right hand columns indicate the prior distribution, $p(\cdot)$, on each parameter. We only give priors for parameters on which we perform sampling. $\mathcal{U}(a,b)$ indicates a uniform distribution between $a$ and $b$, while $\log\mathcal U(a,b)$ indicates a log-uniform distribution between $a$ and $b$.}
    \label{tab:parameters_and_priors}
\end{table*}
\egroup

\subsection{Broad parameter space}
\label{ssec:results_full_parameter_space}
In this section, we consider a suite of 200 simulations whose injected parameter values are drawn from the prior distributions in the far right hand column of Table~\ref{tab:parameters_and_priors}. We use these recoveries to generate PP plots for each parameter. A point on a PP plot indicates the fraction of the 200 injected values that are encapsulated within a confidence interval (vertical axis) versus the confidence interval itself (horizontal axis). Ideally, the PP plot should show a diagonal line for each parameter. We discuss interpretation of PP plots in Appendix~\ref{app:understanding_pp_plots}.

\begin{figure*}
    \centering
    \includegraphics[width=0.49\textwidth]{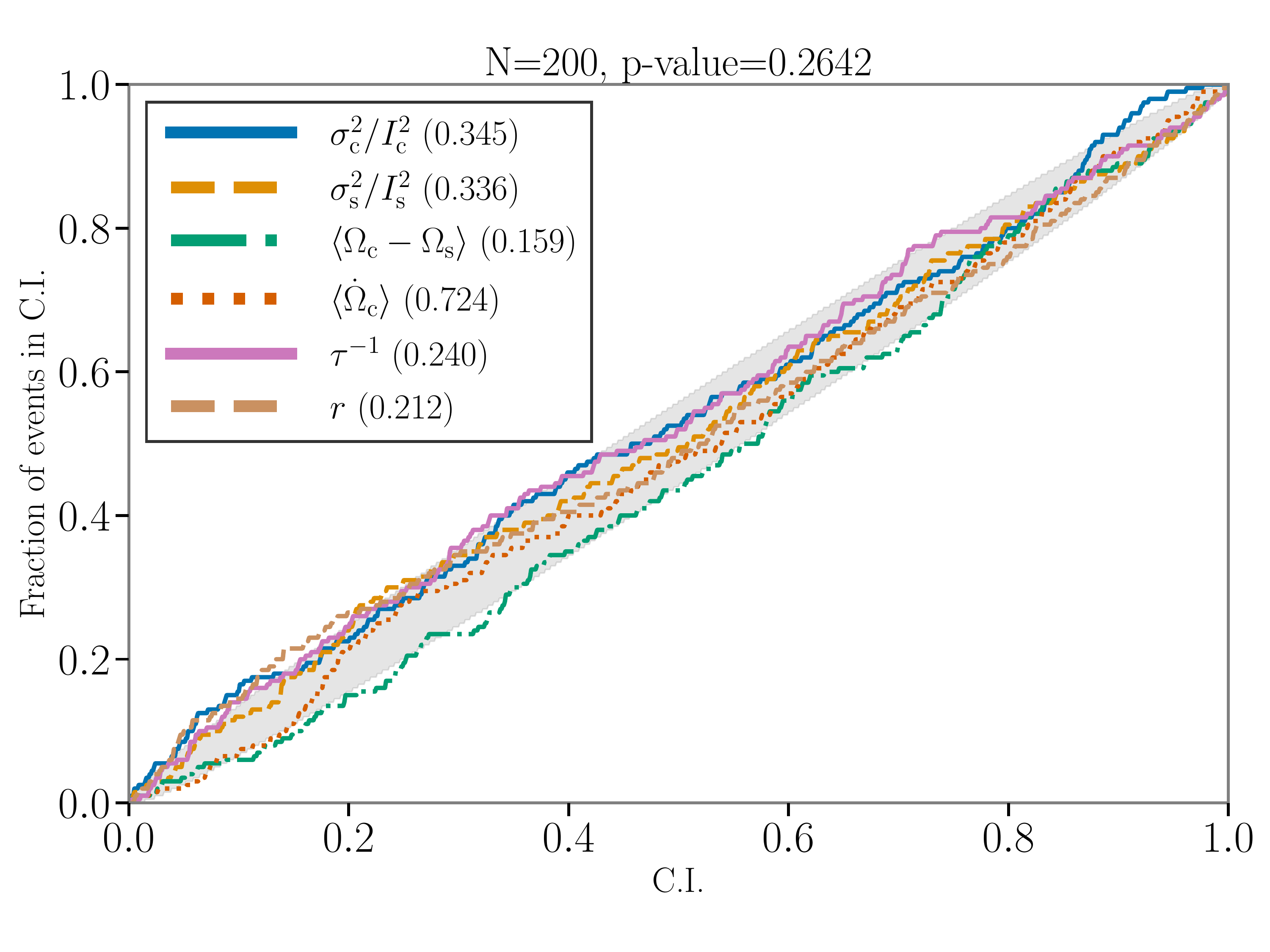}
    \includegraphics[width=0.49\textwidth]{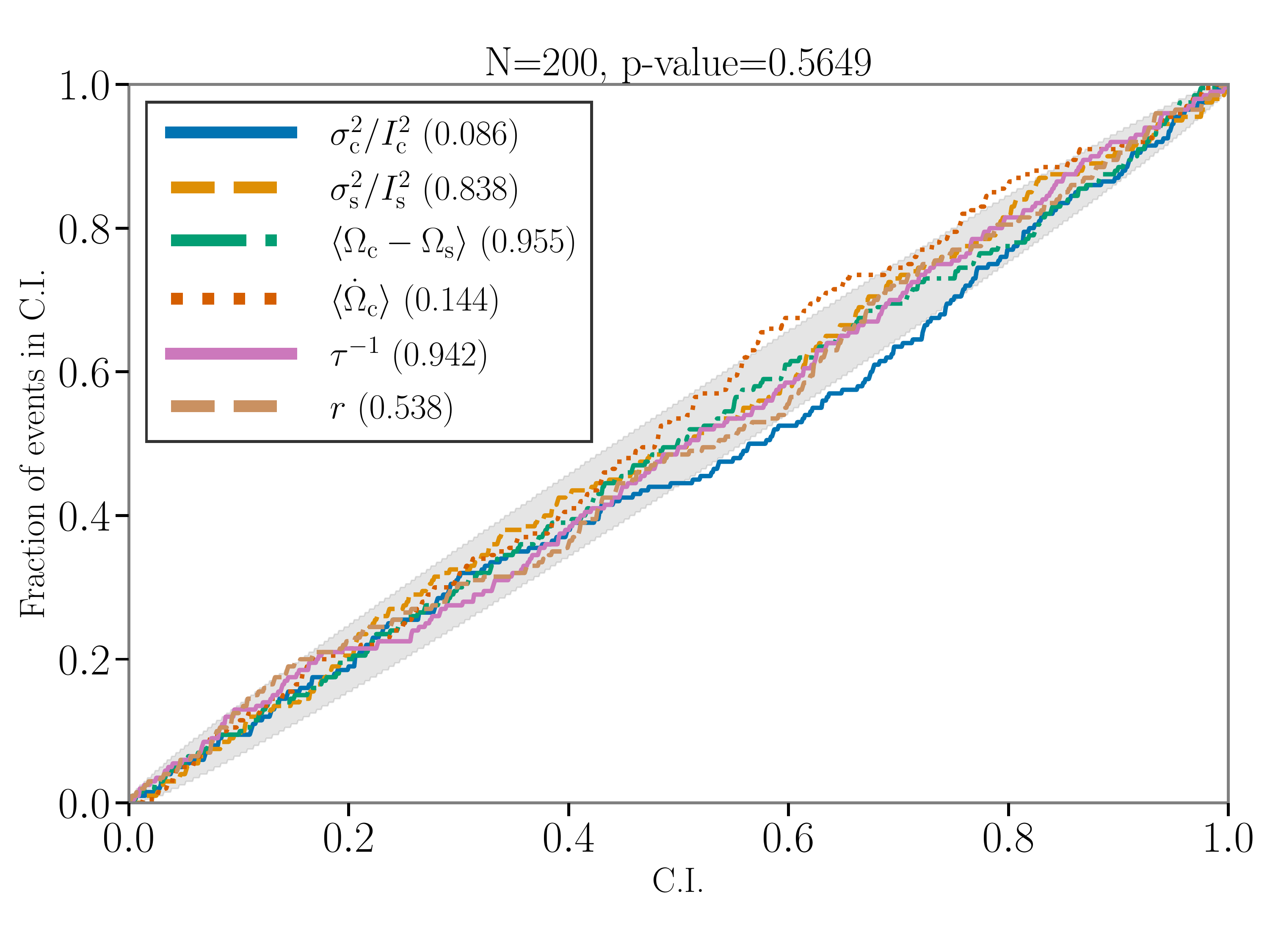}
    \caption{PP plots for future scenario. Each colored curve corresponds to one parameter (color code in the legend). It indicates the fraction of 200 simulated events events whose injected parameter value falls within a certain confidence interval (label C.I.) as a function of that confidence interval. If the posterior distributions are properly estimated, then the curves should fall within the shaded region. The $p$-values in the legend indicate whether the posteriors associated with each parameter are well-behaved (a more technical discussion is given in the text). {\bf Left}: Future scenario with electromagnetic and gravitational-wave measurements for $N_t=600$ spread over 1825 days; {\bf Right}: same as for top left but with $N_t=1200$ spread over 3650 days. }
    \label{fig:pp_plot_em_gw}
\end{figure*}

First, we consider the PP plots for the future electromagnetic and gravitational-wave measurement scenario, which are shown in Fig.~\ref{fig:pp_plot_em_gw}. The left panel uses $N_t = 600$ while the right panel uses $N_t=1200$. The curves for all parameters show the expected linear behavior discussed previously. The number in parentheses next to each parameter gives a $p$-value for the Kolmogorov--Smirnoff test discussed in Appendix~\ref{app:understanding_pp_plots}~\citep{kolmogorov_original,JSSv008i18}. Each of the parameters give $p\gtrsim0.1$, indicating the parameters are drawn from the expected distribution. The shaded region gives 90\% confidence intervals on the excursion one might expect based on the number of simulations performed. Clearly most parameters remain inside this shaded region as well.

Next, we consider the PP plots for the present-day electromagnetic-only measurement scenario, which are shown in the bottom row of Fig.~\ref{fig:pp_plot_em_gw}. The left panel is for $N_t=600$ and the right panel is for $N_t=1200$. Once again, all of the parameters pass the Kolmogorov--Smirnoff test, indicating that our method does a good job of accurately estimating the posterior distribution. 

The results in this section show that over the range of parameters presented in the far right hand column of Table~\ref{tab:parameters_and_priors}, our method produces posterior distributions that are unbiased and accurately reflect our ability to constrain the injected parameters given the data. This test does not make any statements about our ability to detect the relaxation time, $\tau$, over the parameter domain in Table~\ref{tab:parameters_and_priors} in a Bayesian sense. A systematic study of our ability to distinguish between a model that includes the relaxation process and one that does not, over the full parameter domain, is reserved for future work. 
\subsection{Future plans for validation}
\label{ssec:future_plans_for_validation}
We validate the algorithm in this paper by generating data from the model in equations (\ref{eq:crust_equation_of_motion}--\ref{eq:white_noise_torque_covariance}) underpinning the Kalman filter. The model captures certain phenomenological properties of neutron star rotation, specified in Section~\ref{ssec:two_component_model:equations_of_motion}, which have been observed in many pulsars over decades. However, the properties are not universal; some pulsars do not exhibit them at all, while other pulsars exhibit some of them some of the time but not always. For example, the timing noise model discussed in detail in Appendix~\ref{app:power_spectral_analysis} does not apply perfectly to every pulsar; the true power spectrum may be shallower or steeper than $\Phi(f) \propto f^{-4}$. There is also evidence for a cut-off in the power spectrum at low frequencies in some pulsars \citep{2020MNRAS.497.3264G}, which requires extending equations (\ref{eq:crust_equation_of_motion}--\ref{eq:white_noise_torque_covariance}) with an additional filter and hence additional parameters.

A systematic study of whether an unrealistic or simplified noise model leads to systematic biases in the recovery of other physical parameters, like $\tau$, is a subject of ongoing work outside the scope of this paper. In this regard, one must include the choice of priors as part of the model. As a simple example, consider a situation where we analyze a pulsar with very little timing noise, but the prior probability distribution on the noise amplitudes, $\sigmac/\Ic$ and $\sigmas/\Is$, cuts off above the true value for the underlying physical process. It stands to reason that the nested sampling might converge to a value of the relaxation time-scale that is quite short, because the `extra' noise built into the model through our choice of prior could be damped by the relaxation process, resulting in less observed timing variability in $\Omega_{\rm c}$ at the observation epochs. How the observed variability of the crust rotation frequency, the relaxation time-scale, and the white noise amplitudes $\sigmac/\Ic$ and $\sigmas/\Is$, relate to one another is discussed in Appendix D of \citep{exp_max_paper_meyers_melatos}.

\section{Conclusion}
In this paper we develop and characterize a method for analyzing multi-messenger data to estimate the posterior distributions of parameters in the classic crust-superfluid model of a neutron star interior. The method builds on previous work, which used a maximum-likelihood estimator of the two-component model~\citep{exp_max_paper_meyers_melatos}. The current paper extends the maximum-likelihood estimator to compute full posterior distributions for parameters. It also solves the full dynamical system (as opposed to making an Euler approximation) and accommodates non-uniform sampling of the measurements, which is the norm in pulsar timing experiments.

We introduce a Bayesian parameter-estimation framework to estimate the posterior distribution of each parameter in the system. We discuss the Kalman filter used to track the frequency of each component of the star, present its associated log-likelihood function, and discuss how we can perform parameter estimation using MCMC or nested sampling techniques.  We discuss the range of values we expect those parameters to take based on the astrophysical literature.

Finally, we test our method on synthetic data. We first focus on a simple example of an isolated neutron star. We consider cases of $N_t=600$ and $N_t=1200$ spread over 5 and 10 years respectively (and include a third example of $N_t=600$ spread over 20 years in Appendix~\ref{app:less_frequent_observations}). In the future scenario when electromagnetic and gravitational-wave measurements are available, we are able to accurately estimate all of the model parameters, including the noise amplitudes. In the present-day scenario, when only electromagnetic data are available, we estimate $\tau$ with 20\% error and $\langle\dot\Omega_{\rm c}\rangle$ with 4\% error when $N_t=1200$\footnote{Percent errors here represent the percent error of a recovery that peaks one standard deviation from the injected value, with the standard deviation estimated from the width of the posteriors in Fig.~\ref{fig:posteriors_em_gw_simple_example}. These percent errors are parameter dependent. For example, reducing the amount of timing noise by lowering $\sigmac/\Ic$ will result in a more precise measurement of $\langle\dot\Omega_{\rm c}\rangle$.}. This is consistent with the identifiability analysis presented in Section~\ref{ssec:two_component_model:identifiability}, which is carried out under the simplifying assumption of zero noise. We also constrain the ratio of relaxation times, $r$, to within an order of magnitude. In terms of the noise parameters, we estimate $\sigmac^2/\Ic^2$ with 5\% error, and for $N_t=1200$ there is a peak in the $\sigmas^2/\Is^2$ posterior near the injected value in  Fig.~\ref{fig:posteriors_em_only_simple_example}.
The method works reliably across the astrophysically plausible parameter space. Validation tests with 200 randomly-sampled parameter vectors result in reliable posterior distributions that accurately contain the injected parameter vectors, which indicates that the percent errors cited above are consistent with statistical fluctuations. This test confirms that the method can be applied to a wide variety of pulsars without hand-tuning.

The method is now ready to be used on real data, and the code is publicly available\footnote{\url{http://www.github.com/meyers-academic/baboo}}. Future work will focus on generating a time-ordered set of frequency measurements from a set of pulse times-of-arrival~\citep{2018MNRAS.478.3832S,2019MNRAS.485....2C} and running the method on existing and upcoming data sets like UTMOST~\citep{2017PASA...34...45B},  MeerKAT~\citep{2020PASA...37...28B}, and Parkes~\citep{2020PASA...37...20K}. This new method can be added to the list of recent innovations and results in analyzing timing noise~\citep{2019MNRAS.487.5854N,2019MNRAS.489.3810P,parthasarathy:2020wel,2020MNRAS.494..228L,2020MNRAS.497.3264G}, pulsar glitch analysis~\citep{2019NatAs...3.1143A} and pulsar glitch detection schemes built on similar methods~\citep{2020ApJ...896...78M}. 
\label{sec:conclusion}

\section*{Acknowledgements}
The authors acknowledge useful discussions with Sofia Suvorova and William Moran, and Liam Dunn for discussions on integrating the equations of motion. We also thank the insightful anonymous referee. Parts of this research were conducted by the Australian Research Council Centre of Excellence for Gravitational Wave Discovery (OzGrav), through project number CE170100004.

\section*{Data Availability}
No new data were generated or analysed in support of this research.

\appendix
\section{Analytic derivation of power spectrum}
\label{app:power_spectral_analysis}

In this section we derive the timing noise power spectrum of the two-component model given by the differential equations (\ref{eq:crust_equation_of_motion}) and (\ref{eq:superfluid_equation_of_motion}). The constant torques $\Nc$ and $N_s$ do not contribute anything to the power spectrum of the stochastic parts of the solutions so they are removed from the differential equations for this calculation. This is equivalent to subtracting away a linear best fit model. 

A Fourier transformation (denoted by a hat) of the equations (\ref{eq:crust_equation_of_motion}) and (\ref{eq:superfluid_equation_of_motion}) yields
\begin{align}
    i \omega \hat{\Omega}_{\rm c}(\omega) &= - \frac{1}{\tauc} \hat{\Omega}_{\rm c}(\omega) + \frac{1}{\tau_c} \hat{\Omega}_{\rm s}(\omega) + \frac{\hat{\xi}_{\rm c}(\omega)}{\Ic},\\
    i \omega \hat{\Omega}_{\rm s}(\omega) &= - \frac{1}{\taus} \hat{\Omega}_{\rm s}(\omega) + \frac{1}{\taus} \hat{\Omega}_{\rm c}(\omega) + \frac{\hat{\xi}_{\rm s}(\omega)}{\Is}.
\end{align}
Solving these linear equations for $ \hat{\Omega}_{\rm c}(\omega)$ and $ \hat{\Omega}_{\rm s}(\omega)$ gives
\begin{align}
\hat{\Omega}_{\rm c}(\omega) &= \frac{\left( i \omega +\frac{1}{\taus}\right) \frac{\hat{\xi}_{\rm c}(\omega)}{\Ic} + \frac{1}{\tauc}\frac{\hat{\xi}_{\rm s}(\omega)}{\Is}}{-\omega^2 + i \omega/\tau}, \label{eq:omegac_fourier_transform}\\
\hat{\Omega}_{\rm s}(\omega) &= \frac{\frac{1}{\taus}\frac{\hat{\xi}_c(\omega)}{\Ic} + \left( i \omega +\frac{1}{\tauc}\right) \frac{\hat{\xi}_{\rm s}(\omega)}{\Is}}{-\omega^2 + i \omega/\tau}. \label{eq:omegas_fourier_transform}
\end{align}
The power spectra of the stochastic torques can then be inserted to find the power spectra of the angular frequencies. In this paper it is assumed that the torques, $\xi_{\rm c}/\Ic$ and $\xi_{\rm s}/\Is$, are uncorrelated white noise processes with the flat power spectra
\begin{align}
\label{eq:app:xic_assumption}\langle |\hat{\xi}_{\rm c}(\omega)|^2 \rangle &= \sigma_{\rm c}^2,\\
\label{eq:app:xis_assumption}\langle |\hat{\xi}_{\rm s}(\omega)|^2 \rangle &= \sigma_{\rm s}^2,\\
\label{eq:app:xic_xis_assumption}\langle \hat{\xi}^{*}_{\rm c}(\omega) \hat{\xi}_{\rm s}(\omega) \rangle &= 0.
\end{align}
Combining equations (\ref{eq:app:xic_assumption})--(\ref{eq:app:xic_xis_assumption}) with equations (\ref{eq:omegac_fourier_transform}) and (\ref{eq:omegas_fourier_transform}), and applying the Wiener-Khinchin theorem, we obtain
\begin{align}
\label{eq:app:omgc_psd_full}
\langle |\hat{\Omega}_{\rm c}(\omega)|^2 \rangle &= \frac{\left(\omega^2 +\frac{1}{\taus^2}\right) \frac{\sigma^2_{\rm c}}{\Ic^2} + \frac{1}{\tauc^2}\frac{\sigmas^2}{\Is^2}}{\omega^4 + \omega^2/\tau^2}\\
\langle |\hat{\Omega}_{\rm s}(\omega)|^2 \rangle &= \frac{\left(\omega^2 +\frac{1}{\tauc^2}\right) \frac{\sigmas^2}{\Is^2} + \frac{1}{\taus^2}\frac{\sigmac^2}{\Ic^2}}{\omega^4 + \omega^2/\tau^2}\\
\langle \hat{\Omega}^*_{\rm c}(\omega) \hat{\Omega}_{\rm s}(\omega) \rangle &= 
\frac{i \omega \left( \frac{\sigmas^2}{\tauc \Is^2} - \frac{\sigmac^2}{\taus \Ic^2} \right) + \left( \frac{\sigmac^2}{\taus^2 \Ic^2}+  \frac{\sigmas^2}{\tauc^2 \Is^2} \right)}{\omega^4 + \omega^2/\tau^2}.
\end{align}

\begin{figure}
    \centering
    \includegraphics[width=0.4\textwidth]{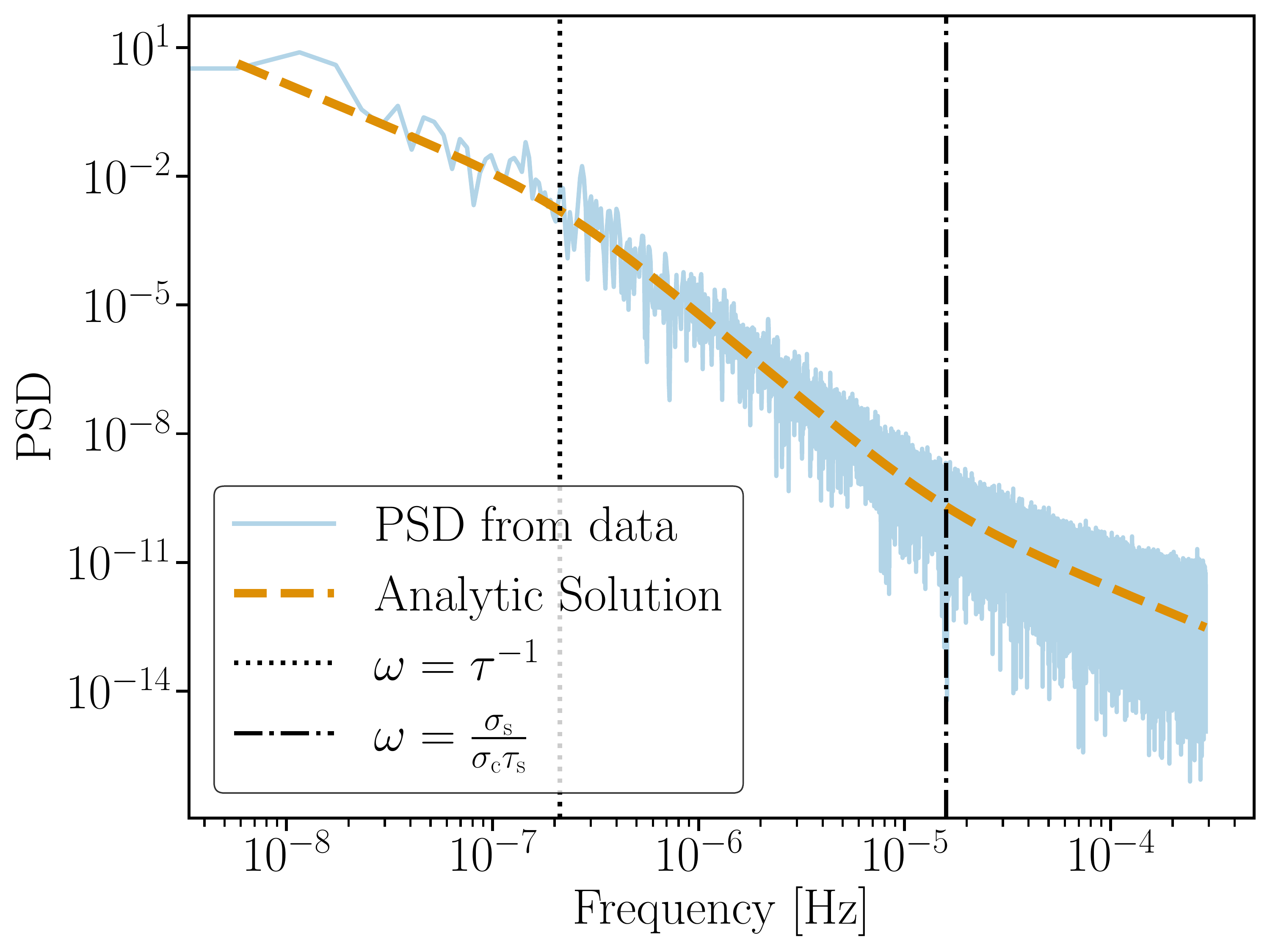}
    \caption{Comparison between simulated PSD of residuals for $\Omegac$ (blue, solid) and analytic solution in equation (\ref{eq:app:omgc_psd_full}) (orange, dashed). The solid blue curve is generated by simulating 2000 days of data sampled 50 times per day. We sample unrealistically often for illustrative purposes. We subtract off a linear fit and take the PSD of the residuals. For this example, we choose $\tauc=10^6~\rm{s}$, $\taus=3\times10^6~\rm{s}$, $\sigmac/\Ic=10^{-9}~\rm{rad~s^{-3/2}}$, $\sigmas/\Is=10^{-7}~\rm{rad~s^{-3/2}}$, indicating that we are in the regime of case I in Table~\ref{tab:spectrum_regimes}. The vertical black (dashed and dash-dotted) lines indicate boundaries between the three sub-cases presented in Table~\ref{tab:spectrum_regimes}.}
    \label{fig:app:psd_plot}
\end{figure}

The power spectra for the angular frequencies $\Omegac$ and $\Omegas$ do not exactly follow power laws. However, in the limits of high and low frequencies the spectra are asymptotic to power laws. To see this, we focus on $\langle|\hat \Omega_{\rm c}|^2\rangle$, which we rewrite as
\begin{align}
\label{eq:app:omgc_psd_full_simplified}
\langle |\hat{\Omega}_{\rm c}(\omega)|^2 \rangle &= \frac{\sigmac^2/\Ic^2}{\omega^2} \frac{\omega^2 +1/\tau'^2}{\omega^2 + 1/\tau^2},
\end{align}
where the timescale $\tau'$ is defined by
\begin{align}
\frac{1}{\tau'^2} = \frac{\sigma^2_{\rm c}/\Ic^2\taus^2 + \sigmas^2/\Is^2\tau_c^2}{\sigmac^2/\Ic^2}.
\end{align}
Depending on which of $\tau$ and $\tau'$ is bigger the power spectrum can have different shapes. In Table~\ref{tab:spectrum_regimes} we summarize the two main cases where one of these timescales dominates the other, along with three sub-cases for each, depending on where in the spectrum we focus.
\begin{table}
\begin{tabular}{c|c|c|c}
\hline  
    && {\bf Case I: } $1/\tau' >1/\tau$ & \\
    \hline
    \hline
    &$\omega < 1/\tau < 1/\tau'$ & $1/\tau < \omega < 1/\tau'$ & $1/\tau < 1/\tau' < \omega$ \\
    \hline
    $|\hat{\Omega}_{\rm c}(\omega)|^2$=&$\frac{\sigmac^2/\Ic^2}{\omega^2} \frac{\tau^2}{\tau'^2}$ & 
    $\frac{\sigmac^2/\Ic^2}{\tau'^2\omega^4}$ & 
    $\frac{\sigmac^2/\Ic^2}{\omega^2}$\\
    \hline
    &&{\bf Case II:} $1/\tau >1/\tau'$&\\
    \hline
    \hline
    &$\omega < 1/\tau' < 1/\tau$ & $1/\tau' < \omega < 1/\tau$ & $1/\tau' < 1/\tau < \omega$ \\
    \hline
    $|\hat{\Omega}_{\rm c}(\omega)|^2$=&$\frac{\sigmac^2/\Ic^2}{\omega^2} \frac{\tau^2}{\tau'^2}$ & 
    $\frac{\sigmac^2}{\Ic^2}\tau^2$ & 
    $\frac{\sigmac^2/\Ic^2}{\omega^2}$\\
    \end{tabular}
    \caption{We show the behaviour of the power spectrum for the crust rotation frequency for two cases, and three regions each. Case I, $1/\tau' > 1/\tau$ is represented in Figure~\ref{fig:app:psd_plot}. The middle region differs between the two cases. In one situation, the power spectrum steepens in this transition region, and in the other case the power spectrum briefly levels off.}
    \label{tab:spectrum_regimes}
\end{table}
The power spectrum for the residual angular velocities generally scales as $\omega^{-2}$, except for a small transition region that scales as $\omega^{-4}$ or as a constant. We show an example in Figure \ref{fig:app:psd_plot} with parameters chosen to accentuate this transition region, and which are consistent with case I in Table~\ref{tab:spectrum_regimes}. In that figure we show representative data (sampled at an unrealistically high rate, and with uniform observation cadence so that we can show the full range of the analytic PSD), for each of the regions we highlight above. We also overlay the full analytic solution in equation~(\ref{eq:app:omgc_psd_full}), and mark each of the limiting cases we discuss by the vertical lines.

In practice, it is common to consider \emph{phase} residuals as opposed to frequency (or angular velocity) residuals. It is straightforward to use the  analytic methods presented in this section to show that the power spectrum for phase residuals is equal to the spectrum for the frequency residuals divided by $\omega^2$. This means that we have a power spectrum in phase residuals that generally goes as $\omega^{-4}$. 

In many papers, e.g.~\cite{2020ApJ...905L..34A}, it is common to take the timing residual power spectral density to be
\begin{align}
P(f) = \frac{A^2}{12\pi^2}\left(\frac{f}{f_{\rm year}}\right)^{-\gamma}f_{\rm year}^{-3}.
\end{align}
If we compare this to the final column of Table~\ref{tab:spectrum_regimes}, then it is straightforward to convert between our model, and typical power-law models in the literature. In our case we have
\begin{align}
\gamma&=4\\
A^2 &= 12\pi^2\frac{\sigmac^2/\Ic^2}{(2\pi)^4f_{\rm rot}^2}f_{\rm year}^{-1},
\end{align}
where $f_{\rm rot}$ is the rotation frequency of the star and is needed to convert between phase residuals (which is how we formulate the problem) and timing residuals (which is generally how the problem is characterized).
\section{Full State Space Representation}
\label{app:state_space_forms}
In this section we give the full forms of the Kalman filter ingredients $\bm F_i$, $\bm T_i$, and $\bm Q_i$ presented in Section~\ref{ssec:two_component_model:state_space_representation} in equations (\ref{eq:transition_integral})--(\ref{eq:process_noise_covariance}).
\begin{align}
\bm F_i &=  \frac{1}{\taus + \tauc}\begin{pmatrix} \tauc+\taus e^{-\Delta t_i/\tau}&\taus - \taus e^{- \Delta t_i/\tau}\\ \tauc-\tauc e^{-\Delta t_i/\tau}&\taus + \tau_c e^{-\Delta t_i/\tau}\end{pmatrix}\label{eq:transition_matrix_full}\\
\bm T_i &=  \begin{bmatrix}\langle \dot\Omega_{\rm c}\rangle \Delta t_i \\
\langle \dot\Omega_{\rm c}\rangle \Delta t_i
\end{bmatrix} +
\tau^2\begin{bmatrix}
\frac{1}{\tauc} \left(\frac{\Nc}{\Ic} -\frac{\Ns}{\Is} \right)\left(1 - e^{-\Delta t_i/\tau}\right)\\
 \frac{1}{\taus}\left(\frac{\Ns}{\Is} - 
\frac{\Nc}{\Ic}\right)\left(1 - e^{-\Delta t_i/\tau}\right)
   \end{bmatrix}
   \label{eq:torque_vector_full}\\
\label{eq:process_noise_covariance_full}
\bm Q_i &= \left(\frac{1}{\tau_c + \tau_s}\right)^2
\begin{pmatrix}
a & b\\
c & d
\end{pmatrix}
\end{align}
with
\begin{align}
\nonumber a=&\Delta t_i\left(\sigcic\tauc^2 + \sigsis\taus^2\right)\\
\nonumber&+ \tau \left(2\sigcic\tauc\taus - 2\sigsis\taus^2\right)\left(1 - e^{-\Delta t_i/\tau}\right)\\
 &+\frac{\tau\taus^2}{2}\left(\sigcic + \sigsis\right)\left(1 - e^{-2\Delta t_i/\tau}\right)\\
\nonumber b=&\Delta t_i\left(\sigcic \tauc^2 + \sigsis\taus^2\right)\\
\nonumber&+ \tau\left(\sigcic\tauc\taus-\tauc^2\sigcic+\sigsis\tauc\taus - \sigsis\taus^2\right)\left(1 - e^{-\Delta t_i/\tau}\right)\\
&-\frac{\tau\tauc\taus}{2}\left(\sigsis + \sigcic\right)\left(1 - e^{-2\Delta t_i/\tau}\right)\\
c=&b\\
\nonumber d=&\Delta t_i\left(\sigcic\tauc^2 + \sigsis\taus^2\right)\\
\nonumber&+ \tau \left(2\sigsis\tauc\taus - 2\sigcic\tauc^2\right)\left(1 - e^{-\Delta t_i/\tau}\right)\\
&+\frac{\tau\tauc^2}{2}\left(\sigcic + \sigsis\right)\left(1 - e^{-2\Delta t_i/\tau}\right).
\end{align}
\section{Kalman filter recursions}
\label{app:kalman_recursions}
We present an overview of the Kalman filter in Section~\ref{sec:parameter_estimation}. In this appendix we discuss practical implementation of the filter.

The Kalman filter is best thought of as taking place over two stages: ``state prediction'' and ``state update.'' In the first step, we predict the current state and its covariance using our estimate of the previous state. In the second step, we use our measurement to update the estimate of the current state. We use the notation $\bm{\hat X}_{i|i-1}$ to denote the estimate of the state $\bm X$ at step $i$ given measurements at steps $1, 2, \ldots, i-1$.  We denote the covariance of the state estimate as $\bm P_{i|i-1} = E[(\bm X_i - \bm{\hat X}_{i|i-1})(\bm X_i - \bm{\hat X}_{i|i-1})^T]$.

The state prediction step is given by using the transition matrix $\bm F_{i}$ to update the state and its covariance
\begin{align}
    \bm{\hat X}_{i | i-1} &= \bm{F}_{i-1}\bm{\hat X}_{i-1|i-1} + \bm T_{i-1}\\
    \bm{P}_{i|i-1} &= \bm{F}_{i-1}\bm P_{i-1|i-1}\bm{F}_{i-1}^T + \bm Q_{i-1}.
\end{align}
The state measurement step then uses the measurement at $t_i$ to update $\bm{\hat X}_i$:
\begin{align}
\bm\epsilon_i &= \bm{Y}_i - \bm C\bm{\hat X}_{i|i-1}\\
\bm S_{i} &= \bm{C}\bm{P}_{i|i-1}\bm{C}^T + \bm R\\
\bm K_{i} &=  \bm{P}_{i|i-1}\bm{C}^T\bm S_{i}^{-1}\\
\bm{\hat{X}}_{i|i} &= \bm{\hat{X}}_{i|i-1} + \bm{K_i}\bm{\epsilon}_i\\
\bm{P}_{i|i} &= \left(\mathbb{I} - \bm{K}_i\bm C\right)\bm{P}_{i|i-1}.
\end{align}
As discussed in Section~\ref{sec:parameter_estimation}, $\bm\epsilon_i$ is typically referred to as the ``innovation,'' and $\bm S_{i}$ is the covariance of the innovation. $\bm K_i$ is known as the ``Kalman gain,''  which is defined so as to minimize $|\bm X_i - \bm{\hat{X}}_{i|i}|^2$, which is equivalent to minimizing the trace of $\bm P_{i|i}$.
 
\section{Kalman Filter likelihood}
\label{app:kf_log_likelihood}
In this appendix we derive the likelihood associated with the Kalman filter that is defined in equation~(\ref{eq:kalman_likelihood}). We begin by noting that for a set of measurements $\{\bm Y_i\} = Y_1, \ldots Y_{N_t}$, the likelihood, conditional on the parameters $\bm\theta$ can be factorized using the chain rule 
\begin{align}
    p(\{\bm Y_i\}|\bm\theta) = \prod_{i=1}^{N_{t}}p(\bm Y_i | \bm Y_{i-1}\ldots Y_1, \bm\theta).
\end{align}
The sub-terms can be further written as a marginalization over the state-variable $\bm X$, given by
\begin{align}
    p(\bm Y_i | \bm Y_{i-1}\ldots Y_1, \bm\theta) = \int \mathrm{d}\bm X_{i}\,p(\bm Y_i | \bm X_i, \bm\theta)p(\bm X_i | \bm Y_{i-1}\ldots \bm Y_1, \bm\theta).
\end{align}
Assuming Gaussian measurement errors and Gaussian errors on the state variables we replace the two probability distributions in the integrand with normal distributions. We use $\mathcal N(x;\mu, \sigma^2)$ to indicate that the random variable $x$ follows a normal distribution with mean $\mu$ and variance $\sigma^2$. We find
\begin{align}
    p(\bm Y_i | \bm Y_{i-1}\ldots Y_1, \bm\theta) =& \int \mathrm{d}\bm X_{i}\,\mathcal N(\bm Y_i ; \bm C\bm X_i,\bm R  |\bm\theta)\nonumber\\
    &\times \mathcal{N}(\bm X_i ; \bm \hat{\bm{X}}_{i|i-1}, \bm P_{i|i-1}| \bm\theta),\label{eq:ll_integral}\\
    =&~\mathcal N(\bm Y_{i} ; \bm C\hat{\bm{X}}_{i|i-1}, \bm S_{i}|\bm\theta),\label{eq:ll_final_result}
\end{align}
where equation~(\ref{eq:ll_final_result}) follows from carrying out the Gaussian integral in equation~(\ref{eq:ll_integral}).
This leads us to the final form of the likelihood
\begin{align}
    p(\{\bm Y_i\}|\bm\theta) = \prod_{i=1}^{N_{t}}\mathcal N(\bm Y_{i} ; \bm C\hat{\bm{X}}_{i|i-1}, \bm S_{i}|\bm\theta),
\end{align}
which is equivalent to equation~(\ref{eq:kalman_likelihood}).
In practice we work with $\log p(\{\bm Y_i\}|\bm\theta)$, which we can calculate recursively during each Kalman filter state update.

\section{Results with less frequent observations}
\label{app:less_frequent_observations}
The frequency with which pulsars are observed depends on the telescope. Telescopes that use non-steerable parabolic reflectors and use interferometry to reconstruct a source on the sky, like the UTMOST project~\citep{2017PASA...34...45B} and the Canadian Hydrogen Intensity Mapping Experiment~\citep{10.1117/12.2054950}, are capable of producing timing results each sidereal day as the source transits across their field of view. Meanwhile, telescopes or arrays consisting of fully-steerable dishes, such as the Green Bank Telescope, The Jodrell Bank Telescope, MeerKAT ~\citep{2020PASA...37...28B}, and the Parkes Radio Telescope~\citep{2020PASA...37...20K} (among many others) must be pointed directly at a source. Such telescopes have the advantage of being more sensitive, but tend to time individual pulsars once per week to once per month. In this scenario, 600 observations over 5 years, or 1200 observations over 10 years, which we use in Sections~\ref{ssec:results_on_a_simple_example} and~\ref{ssec:results_full_parameter_space}, are too closely spaced together to be realistic. Here, we consider a scenario where we have 600 observations spaced over 20 years (and also 1200 observations spaced over 20 years).

\begin{figure*}
    \centering
\includegraphics[width=0.89\textwidth]{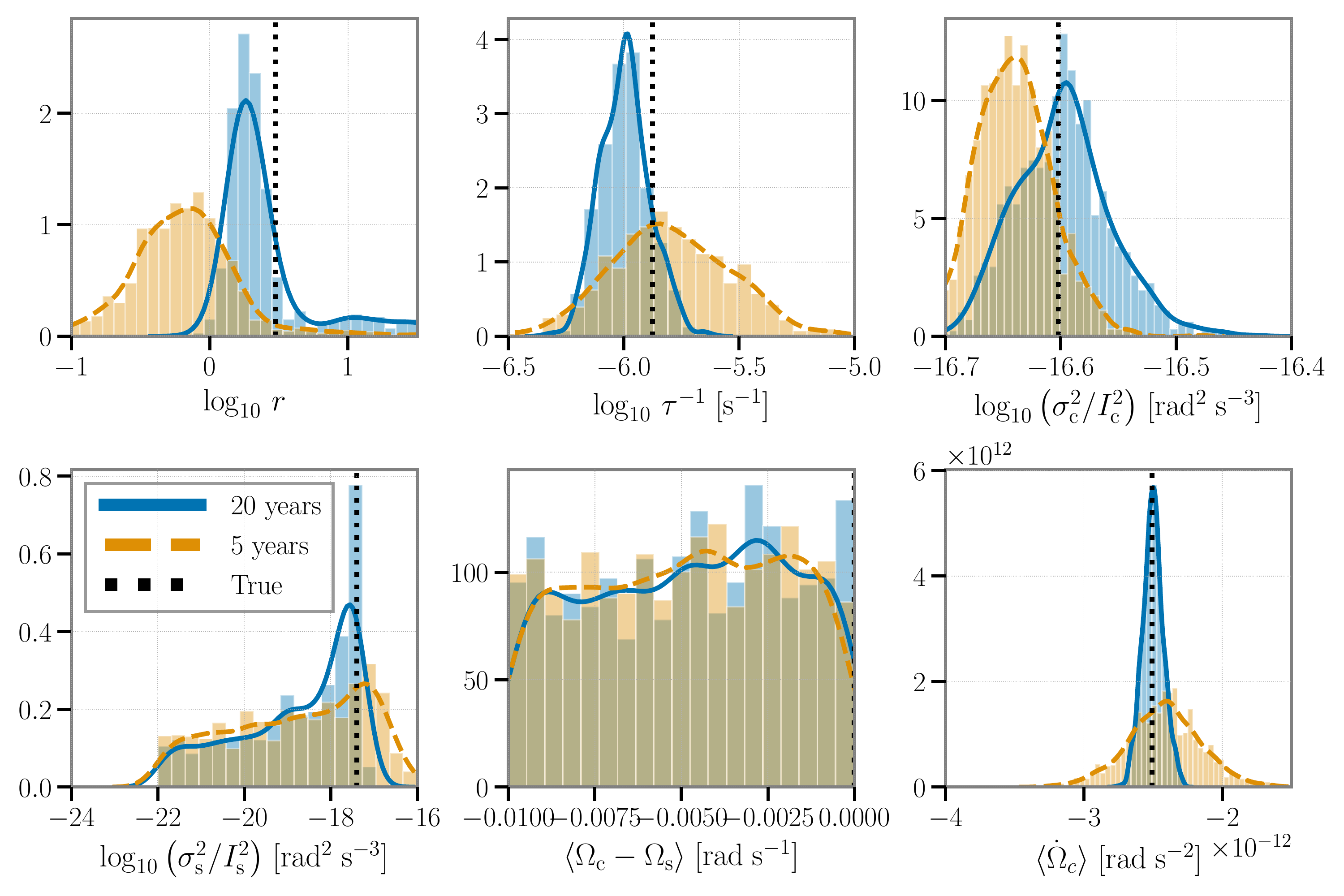}
    
    \caption{1D posteriors for $N_t=600$ for 20 years (blue, solid) and 5 years (orange, dashed), for the present-day observational scenario. The black dotted lines indicate the injected values, which can be found in Table~\ref{tab:parameters_and_priors}. We see that these results broadly agree with the posteriors in Figure~\ref{fig:posteriors_em_only_simple_example}. As one might expect, when the observations are spread over a longer period of time, the posterior on $\langle \dot{\Omega_{\rm c}}\rangle$ narrows.}
    \label{fig:1d_posteriors_long}
\end{figure*}

In Figure~\ref{fig:1d_posteriors_long}, we show 1-dimensional posteriors on the six parameters we attempt to recover using electromagnetic-only observations, where the injection parameters are from Table~\ref{tab:parameters_and_priors}. We show results for $N_t=600$ measurements over 20 years (solid, blue) and over 5 years (orange, dashed). We see that, for the set of parameters used in Table~\ref{tab:parameters_and_priors} and Figure~\ref{fig:posteriors_em_only_simple_example} we are still able to accurately recover the same sets of parameters. For $\langle\dot{\Omega}_{\rm c}\rangle$, the posterior narrows when the observations are spaced over a longer period of time, as one would expect. We also see the posterior on $\tau^{-1}$ narrow.

\begin{figure}
    \centering
    \includegraphics[width=0.49\textwidth]{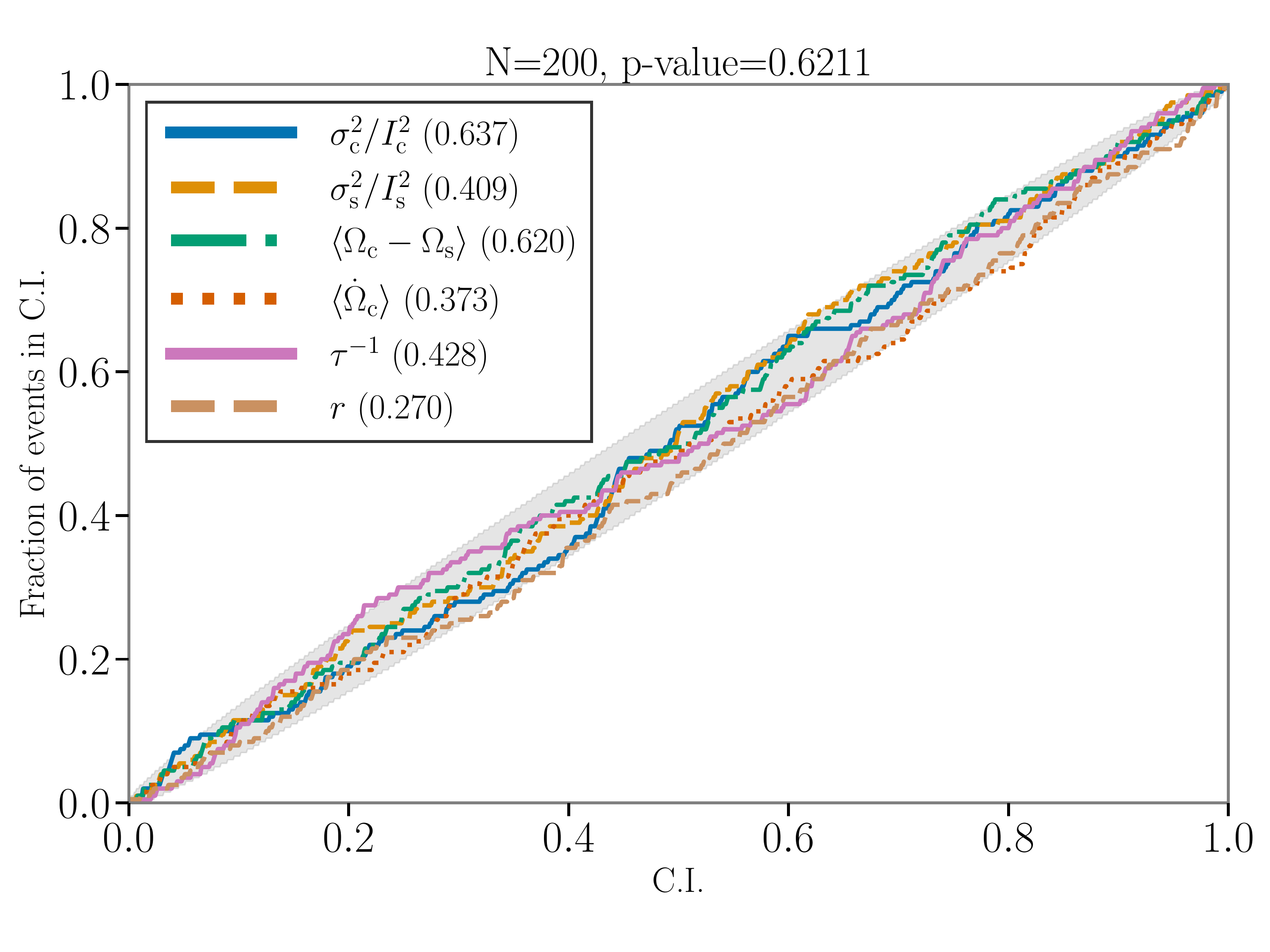}
    
    \caption{PP plots for present-day observational scenario with $N_t=600$, spread over 7300 days (20 years). Each colored curve corresponds to one parameter (color code in the legend). It indicates the fraction of 200 simulated events events whose injected parameter value falls within a certain confidence interval (label C.I.) as a function of that confidence interval. If the posterior distributions are properly estimated, then the curves should fall within the shaded region. The $p$-values in the legend indicate whether the posteriors associated with each parameter are well-behaved (a more technical discussion is given in the text).}
    \label{fig:pp_plot_em_long}
\end{figure}

In Figure~\ref{fig:pp_plot_em_long}, we show PP plots for all six parameters for the $N_t=600$ measurements over 20 years scenario, evaluated over the prior range described in Table~\ref{tab:parameters_and_priors} column 5. Again, the PP plots indicate that we are correctly estimating the posterior of each parameter. It is important to note that, while we are able to accurately evaluate the posteriors over this range, we are not making any claims about our ability to measure a relaxation time-scale, or the width of the posteriors beyond the fact that they are internally consistent. As stated in Sections~\ref{ssec:results_full_parameter_space} and~\ref{ssec:future_plans_for_validation}, such a study is reserved for future work.


\section{Interpretation of PP plots}
\label{app:understanding_pp_plots}
To understand a PP plot intuitively, consider injection $j$ (where $j$ labels injection number and runs from 1 to 200), with injected moment of inertia ratio, $r_j$, for example. We define the cumulative distribution function of the one-dimensional marginalized posterior, 
\begin{align}
F_j(x) = \int_{-\infty}^{x}\textrm{d}r\,p(r|\{\bm Y\}).
\end{align}
If the estimate of $p(r|\{\bm Y\})$, from the nested sampler (see, e.g. the top panel of the left-most column of Fig.~\ref{fig:posteriors_em_gw_simple_example}), is close to the true posterior distribution, then $F_j(r_j)$ is drawn from a uniform distribution. We can test whether our sampler is producing reasonable posterior distributions of $r_j$ by comparing the set of $F_j(r_j)$ values for $j=1\ldots 200$ to a uniform distribution using a Kolmogorov--Smirnoff test.


\bsp	
\label{lastpage}
\end{document}